\documentclass[10pt,conference]{IEEEtran}

\renewcommand\thesection{\arabic{section}}
\renewcommand\thesubsection{\thesection.\arabic{subsection}}
\renewcommand\thesubsubsection{\thesubsection.\arabic{subsubsection}}

\renewcommand\thesectiondis{\arabic{section}}
\renewcommand\thesubsectiondis{\thesectiondis.\arabic{subsection}}
\renewcommand\thesubsubsectiondis{\thesubsectiondis.\arabic{subsubsection}}

\usepackage{verbatim} 
\usepackage[pdftex]{graphicx}
\usepackage{setspace}
\usepackage[cmex10]{amsmath}
\usepackage{amsmath, amsthm, amssymb}
\usepackage{algorithmic}
\usepackage{algorithm}
\usepackage{caption}
\usepackage{subcaption}
\usepackage{paralist}
\usepackage{cases}
\usepackage{cite}

\usepackage{tikz}

\usepackage{color}

\newcommand{\ceiling}[1]{\left\lceil{#1}\right\rceil}

\newcommand{\setof}[1]{\left\{{#1}\right\}}

\newcommand{\toolbox}[1]{{\bf CHL}}
\newcommand{\frameworkkq}[1]{$\mathbf{k^2Q}$}
\newcommand{\frameworkku}[1]{$\mathbf{k^2U}$}

 \def\myendproof{{\ \vbox{\hrule\hbox{%
   \vrule height1.3ex\hskip0.8ex\vrule}\hrule }}\par}
 
 \setboolean{ALC@noend}{true}

\newtheorem{theorem}{Theorem}
\newtheorem{lemma}{Lemma}
\newtheorem{corollary}{Corollary}

\newtheorem{definition}{Definition}

\graphicspath{{fig/multiframe/}{fig/arbitrary/}{fig/dag/}} 

\usetikzlibrary{%
  arrows,%
  shapes.misc,
  shapes.arrows,%
  chains,%
  matrix,%
  positioning,
  scopes,%
  decorations.pathmorphing,
  shadows%
}

\pgfdeclarelayer{background}
\pgfdeclarelayer{foreground}
\pgfsetlayers{background,main,foreground}
\tikzstyle{materia}=[draw, fill=white, text width=1.0em, text centered,
  minimum height=4em,drop shadow]
\tikzstyle{practica} = [materia, text width=18em, minimum width=8em,
align =left,
  minimum height=3em, rounded corners, drop shadow]
\tikzstyle{texto} = [above, text width=6em, text centered]
\tikzstyle{linepart} = [draw, thick, color=blue!50, -latex', dashed]
\tikzstyle{line} = [draw, line width = 2pt, color=blue!50, -latex']
\tikzstyle{ur}=[draw, text centered, minimum height=0.01em]

\newcommand{\practica}[2]{node (p#1) [practica]
  {\\{\footnotesize{#2}}}}

\title{\huge Evaluate and Compare Two Utilization-Based
  Schedulability-Test Frameworks for Real-Time Systems}

\author{
    Jian-Jia Chen and Wen-Hung Huang\\
    Department of Informatics\\
    TU Dortmund University, Germany
    \and
    Cong Liu\\
    Department of Computer Science\\
    The University of Texas at Dallas
}
\vspace{3mm}

\begin{document}

\maketitle
\thispagestyle{plain}

\begin{abstract}
  This report summarizes two general frameworks, namely \frameworkkq{}
  and \frameworkku{}, that have been recently developed by us. The
  purpose of this report is to provide detailed evaluations and
  comparisons of these two frameworks. These two frameworks share some
  similar characteristics, but they are useful for different
  application cases.  These two frameworks together provide
  comprehensive means for the users to automatically convert the
  pseudo polynomial-time tests (or even exponential-time tests) into
  polynomial-time tests with closed mathematical forms.  With the
  quadratic and hyperbolic forms, \frameworkkq{} and \frameworkku{}
  frameworks can be used to provide many quantitive features to be
  measured and evaluated, like the total utilization bounds, speed-up factors, etc.,
  not only for uniprocessor scheduling but also for multiprocessor
  scheduling.  These frameworks can be viewed as ``blackbox''
  interfaces for providing polynomial-time schedulability tests and response time
  analysis for real-time applications.  We have already presented
  their advantages for being applied in some models in the previous
  papers.  However, it was not possible to present a more
  comprehensive comparison between these two frameworks. We hope this
  report can help the readers and users clearly understand the
  difference of these two frameworks, their unique characteristics,
  and their advantages. We demonstrate their differences and
  properties by using the traditional sporadic real-time task models
  in uniprocessor scheduling and multiprocessor global scheduling.
\end{abstract} 

\section{Introduction}
\label{sec:intro}

To analyze the worst-case response time or to ensure the timeliness of
the system, for each of individual task models, researchers tend to develop
dedicated techniques that result in schedulability tests with
different computation complexity and accuracy of the
analysis. Although many successful results have been developed, after
many real-time systems researchers devoted themselves for many years,
there does not exist a general framework that can provide  efficient and
effective analysis for different task models.

A very widely adopted case is the schedulability test of a
(constrained-deadline) sporadic real-time task $\tau_k$ under
fixed-priority scheduling in uniprocessor systems, in which the
time-demand analysis (TDA) developed in
\cite{DBLP:conf/rtss/LehoczkySD89} can be adopted. That is, if
\begin{equation}
  \label{eq:exact-test-constrained-deadline}
\exists t \mbox{ with } 0 < t \leq D_k {\;\; and \;\;} C_k +
\sum_{\tau_i \in hp(\tau_k)} \ceiling{\frac{t}{T_i}}C_i \leq t,
\end{equation}
then task $\tau_k$ is schedulable under the fixed-priority scheduling algorithm, where $hp(\tau_k)$ is the set of tasks with higher priority than $\tau_k$, $D_k$, $C_k$, and $T_i$ represent $\tau_k$'s relative deadline, worst-case execution time, and period, respectively. TDA requires pseudo-polynomial-time complexity to check the time points that lie in $(0, D_k]$ for Eq.~\eqref{eq:exact-test-constrained-deadline}. 

However, it is not always necessary to test all possible time points
to derive a safe worst-case response time or to provide sufficient
schedulability tests.
The general and key concept to obtain sufficient schedulability tests in
\frameworkku{} in \cite{DBLP:journals/corr/abs-1501.07084,DBLP:conf/rtss/ChenHL15} and
\frameworkkq{} in \cite{DBLP:journals/corr/abs-kRTA,DBLP:conf/rtss/ChenHL16} is to test only a
subset of such points for verifying the schedulability. 
Traditional fixed-priority schedulability tests often have
pseudo-polynomial-time (or even higher) complexity. 
The idea
implemented in the \frameworkku{} and \frameworkkq{} frameworks  is to 
provide a general $k$-point schedulability test, which
only needs to test $k$ points under \textit{any} fixed-priority
scheduling when checking schedulability of the task with the $k^{th}$
highest priority in the system.  Moreover, this concept is further
extended in \frameworkkq{} to provide a safe upper bound of the
worst-case response time. The response time analysis and the
schedulability analysis provided by the frameworks can be viewed as
``\emph{blackbox}" interfaces that can result in sufficient utilization-based
analysis, in which the utilization of a task is its execution time
divided by its period.

The \frameworkku{} and \frameworkkq{}
frameworks are in fact two different important components for building
efficient and effective schedulability tests and response time
analysis. Even though they come from the same observations by testing
only $k$ effective points, they are in fact fundamentally different
from mathematical formulations and have different properties.  In
\frameworkku{}, all the testings and formulations are based on the
task utilizations.  In \frameworkkq{}, the testings are based not only
on the task utilizations, but also on the task execution times.  The
different formulations of testings result in different types of
solutions.  The natural form of \frameworkku{} is a hyperbolic form
for testing the schedulability of a task, whereas the natural form of
\frameworkkq{} is a quadratic form for testing the schedulability or the
response time. In general, if the $k$ points can be effectively
defined, \frameworkku{} has more precise results. However, if these
$k$ points cannot be easily defined or there is some ambiguity to fine
the effective points, then \frameworkkq{} may be more suitable for such
models.

There have been several results in the literature with respect to
utilization-based, e.g.,
\cite{liu1973scheduling,HanTyan-RTSS97,journals/tc/LeeSP04,DBLP:conf/rtas/WuLZ05,kuo2003efficient}
for the sporadic real-time task model and its generalizations in uniprocessor systems.
The novelty of \frameworkku{} and \frameworkkq{} comes from a different
perspective from these approaches
\cite{liu1973scheduling,HanTyan-RTSS97,journals/tc/LeeSP04,DBLP:conf/rtas/WuLZ05,kuo2003efficient}. We
do not specifically seek for the total utilization bound. Instead, we
look for the critical value in the specified sufficient schedulability
test while verifying the schedulability of task $\tau_k$. 
The natural condition to test
the schedulability of task $\tau_k$ is a hyperbolic bound when
\frameworkku{} is adopted, whereas the nature condition to
test task $\tau_k$ is a quadratic bound when \frameworkkq{} is adopted
(to be shown in Section~\ref{sec:framework}). The corresponding total
utilization bound can be obtained.

The generality of the \frameworkkq{} and \frameworkku{} frameworks has
been demonstrated in
\cite{DBLP:journals/corr/abs-1501.07084,DBLP:conf/rtss/ChenHL15,DBLP:journals/corr/abs-kRTA,DBLP:conf/rtss/ChenHL16}.
We believe that these two frameworks, to be used for different cases,
have great potential in analyzing many other complex real-time task
models, where the existing analysis approaches are insufficient or
cumbersome.  We have already presented their advantages for being
applied in some models in
\cite{DBLP:journals/corr/abs-1501.07084,DBLP:conf/rtss/ChenHL15,DBLP:journals/corr/abs-kRTA,DBLP:conf/rtss/ChenHL16}.
However, it was not possible to present a more comprehensive
comparison between these two frameworks in \cite{DBLP:journals/corr/abs-1501.07084,DBLP:conf/rtss/ChenHL15,DBLP:journals/corr/abs-kRTA,DBLP:conf/rtss/ChenHL16}. We hope this report can help
the readers and users clearly understand the difference of these
two frameworks, their unique characteristics, and their advantages.
Since our focus in this report is only to demonstrate the similarity,
the difference
and the characteristics of these two frameworks, we will use 
the simplest setting, i.e., the traditional sporadic real-time task
models in uniprocessor scheduling and multiprocessor global
scheduling.

For the \frameworkkq{} and \frameworkku{} frameworks, their
characteristics and advantages over other approaches have been already
discussed in
\cite{DBLP:journals/corr/abs-1501.07084,DBLP:conf/rtss/ChenHL15,DBLP:journals/corr/abs-kRTA,DBLP:conf/rtss/ChenHL16}. However,
between these two frameworks, we only gave short sketches and high-level descriptions of their
differences and importance. These explanations may seem
incomplete in \cite{DBLP:journals/corr/abs-1501.07084,DBLP:conf/rtss/ChenHL15,DBLP:journals/corr/abs-kRTA,DBLP:conf/rtss/ChenHL16} to explain whether both are needed or only one of them is important.  Therefore,
we would like to present in this report to explain why both frameworks
are needed and have to be applied for different cases.  Moreover, we
would like to emphasize that both frameworks are
important. In general, the \frameworkku{}
framework is more precise by using only the utilization values of the
higher-priority tasks. If we can formulate the schedulability tests
into the \frameworkku{} framework, it is also usually possible to
model it into the \frameworkkq{} framework. In such cases, the same
pseudo-polynomial-time test is used. When we consider the worst-case
quantitive metrics like utilization bounds, resource augmentation
bounds, or speedup factors, the result derived from the \frameworkku{}
framework is better for such cases. However, there are also cases, in
which formulating the test by using the \frameworkku{} framework is
not possible. These cases may even start from schedulability tests
with exponential-time complexity. We have successfully demonstrated
three examples in \cite{DBLP:journals/corr/abs-kRTA} by using the
\frameworkkq{} framework to derive polynomial-time tests. In those demonstrated cases, either
the \frameworkku{} framework cannot be applied or with worse results
(since different exponential-time or pseudo-polynomial-time schedulability tests are applied).

\noindent\textbf{Organizations.} The rest of this report is organized
as follows:
\begin{compactitem}
\item The basic terminologies and models are presented in
  Section~\ref{sec:model}. 
\item The two frameworks
  from\cite{DBLP:journals/corr/abs-1501.07084,DBLP:conf/rtss/ChenHL15,DBLP:journals/corr/abs-kRTA,DBLP:conf/rtss/ChenHL16}
  are summarized and presented in Section~\ref{sec:framework}.
\item We demonstrate two different comparisons between the frameworks
  by using sporadic task systems in uniprocessor systems and
  multiprocessor systems. 
\end{compactitem}
Note that this report does not intend to provide new theoretical
results. All the omitted proofs are already provided in
\cite{DBLP:journals/corr/abs-1501.07084,DBLP:conf/rtss/ChenHL15,DBLP:journals/corr/abs-kRTA,DBLP:conf/rtss/ChenHL16}. For
some simple properties derived from the results in
\cite{DBLP:journals/corr/abs-1501.07084,DBLP:conf/rtss/ChenHL15,DBLP:journals/corr/abs-kRTA,DBLP:conf/rtss/ChenHL16},
we will explain how such results are derived.


\section{Basic Task and Scheduling Models}
\label{sec:model}

This report will demonstrate the effectiveness and differences of the
two frameworks by using the sporadic real-time task model, even though
the frameworks target at more general task models. We define the
terminologies in this section for completeness.  A sporadic task
$\tau_i$ is released repeatedly, with each such invocation called a
job. The $j^{th}$ job of $\tau_i$, denoted $\tau_{i,j}$, is released
at time $r_{i,j}$ and has an absolute deadline at time $d_{i,j}$. Each
job of any task $\tau_i$ is assumed to have execution time $C_i$. Here
in this report, whenever we refer to the execution time of a job, we
mean for the worst-case execution time of the job, since all the analyses we use are safe by only considering 
the worst-case execution time.  The response time
of a job is defined as its finishing time minus its release
time. Successive jobs of the same task are required to execute in
sequence. Associated with each task $\tau_i$ are a period $T_i$, which
specifies the minimum time between two consecutive job releases of
$\tau_i$, and a deadline $D_i$, which specifies the relative deadline
of each such job, i.e., $d_{i,j}=r_{i,j}+D_i$. The worst-case response
time of a task $\tau_i$ is the maximum response time among all its
jobs.  The utilization of a task $\tau_i$ is defined as $U_i=C_i/T_i$.

A sporadic task system $\tau$ is said to be an implicit-deadline
system if $D_i = T_i$ holds for each $\tau_i$. A sporadic task system
$\tau$ is said to be a constrained-deadline system if $D_i \leq T_i$
holds for each $\tau_i$.  Otherwise, such a sporadic task system
$\tau$ is an arbitrary-deadline system.

A task is said \emph{schedulable} by a scheduling policy if all of its
jobs can finish before their absolute deadlines, i.e., the worst-case
response time of the task is no more than its relative deadline.  A
task system is said \emph{schedulable} by a scheduling policy if all
the tasks in the task system are schedulable. A \emph{schedulability
  test} is to provide sufficient conditions to ensure the feasibility
of the resulting schedule by a scheduling policy. 

Throughout the report, we will focus on fixed-priority preemptive
scheduling. That is, each task is associated with a priority
level. More specifically, we will only use rate monotonic (RM, i.e., 
tasks with smaller periods are with higher priority levels) and 
deadline monotonic (DM, i.e., 
tasks with smaller relative deadlines are with higher priority levels)
in this report.
For a uniprocessor system, the scheduler always dispatches the job
with the highest priority in the ready queue to be executed.  For a
multiprocessor system, we consider multiprocessor global scheduling on
$M$ identical processors, in which each of them has the same
computation power. For global multiprocessor scheduling, there is a
global queue and a global scheduler to dispatch the jobs. We consider
only global fixed-priority scheduling. At any time, the
$M$-highest-priority jobs in the ready queue are dispatched and
executed on these $M$ processors.

Note that the above definitions are just for simplifying the
presentation flow in this report. The frameworks can still work for
non-preemptive scheduling and different types of fixed-priority
scheduling. 

We will only present the schedulability test of a certain task
$\tau_k$, that is being analyzed, under the above assumption. For
notational brevity, in the framework presentation, we will implicitly
assume that there are $k-1$ tasks, says $\tau_1, \tau_2, \ldots,
\tau_{k-1}$ with higher-priority than task $\tau_k$. \emph{These $k-1$
higher-priority tasks are assumed to schedulable before we move on
to test task $\tau_k$.} We will use
$hp(\tau_k)$ to denote the set of these $k-1$ higher priority tasks,
when their orderings do not matter. Moreover, we only
consider the cases when $k \geq 2$, since $k=1$ is pretty trivial.

\section{\frameworkku{} and \frameworkkq{} Frameworks}
\label{sec:framework}

This section presents the definitions and properties of the \frameworkku{} and \frameworkkq{} frameworks
for testing the schedulability of task $\tau_k$ in a given set of
real-time task. The construction of the frameworks requires the
following definitions:

\begin{definition}
  \label{def:kpoints-k2u}
  A $k$-point effective schedulability test is a sufficient schedulability test of a fixed-priority scheduling policy, that verifies the existence of $t_j \in \setof{t_1, t_2, \ldots t_k}$ with $0 < t_1 \leq t_2 \leq \cdots \leq t_k$ such that \begin{equation}
    \label{eq:precodition-schedulability-k2u}
    C_k + \sum_{i=1}^{k-1} \alpha_i t_i U_i + \sum_{i=1}^{j-1} \beta_i t_i U_i \leq t_j,
  \end{equation}
  where $C_k > 0$, $\alpha_i > 0$, $U_i > 0$, and $\beta_i >0$ are dependent upon the setting
  of the task models and task $\tau_i$. \myendproof
\end{definition}

\begin{definition}[Last Release Time Ordering]
Let $\pi$ be the last release time ordering assignment as a bijective
function $\pi:  hp(\tau_k)
\rightarrow \setof{1, 2, \ldots,k-1}$ to define the last release time ordering
of task $\tau_j \in hp(\tau_k)$ in the window of interest. Last release time orderings are
numbered from $1$ to $k-1$, i.e., $|hp(\tau_k)|$, where 1 is the earliest and $k-1$ the
latest. \myendproof
\end{definition}

\begin{definition}
  \label{def:kpoints}
  A $k$-point last-release schedulability test under a given ordering
  $\pi$ of the $k-1$ higher priority tasks is a sufficient schedulability test of a fixed-priority scheduling policy, that verifies the existence of  $0 \leq t_1 \leq t_2 \leq \cdots \leq t_{k-1}  \leq t_k$ such that \begin{equation}
    \label{eq:precodition-schedulability}
    C_k + \sum_{i=1}^{k-1} \alpha_i t_i U_i + \sum_{i=1}^{j-1} \beta_i C_i \leq t_j, 
  \end{equation}
  where $C_k > 0$, for $i=1,2,\ldots,k-1$, $\alpha_i > 0$, $U_i > 0$, $C_i \geq 0$, and $\beta_i >0$ are dependent upon the setting
  of the task models and task $\tau_i$. \myendproof
\end{definition}
 
\begin{definition}
  \label{def:kpoints-response}
  A $k$-point last-release response time analysis is a safe response time
  analysis of a fixed-priority scheduling policy under a given
  ordering $\pi$ of the $k-1$ higher-priority tasks by finding the
  maximum
  \begin{equation}
    \label{eq:precond-objective-k}
   t_k = C_k + \sum_{i=1}^{k-1} \alpha_i t_i U_i + \sum_{i=1}^{k-1} \beta_i C_i,
  \end{equation}
with $0 \leq t_1 \leq t_2 \leq \cdots \leq t_{k-1} \leq t_{k}$ and
  \begin{align}
    \label{eq:precond-objective-k2}
    C_k + \sum_{i=1}^{k-1} \alpha_i t_i U_i + \sum_{i=1}^{j-1} \beta_i C_i > t_j, & \forall j=1,2,\ldots,k-1,
  \end{align}
  where $C_k > 0$, $\alpha_i > 0$, $U_i > 0$, $C_i \geq 0$, and $\beta_i >0$ are dependent upon the setting
  of the task models and task $\tau_i$. \myendproof
\end{definition}

Throughout the report, we implicitly assume that $0 < t_k \neq \infty$ when
Definition~\ref{def:kpoints-k2u} and Definition~\ref{def:kpoints} are
used, as $t_k$ is usually related to the given relative deadline
requirement. Note that $t_k$ may still become $\infty$ when
Definition~\ref{def:kpoints-response} for response time analysis is
used. Moreover, we only consider non-trivial cases, in which $C_k > 0$, and
$\alpha_i > 0$, $\beta_i > 0$, $C_i \geq 0$,
and $0 < U_i \leq 1$ for $i=1,2,\ldots,k-1$.


\subsection{Comparison of Definition~\ref{def:kpoints-k2u} and Definition~\ref{def:kpoints}}

The definition of the $k$-point last-release schedulability test $C_k
+ \sum_{i=1}^{k-1} \alpha_i t_i U_i + \sum_{i=1}^{j-1} \beta_i C_i
\leq t_j$ in Definition \ref{def:kpoints} only slightly differs from
the $k$-point effective schedulability test $C_k + \sum_{i=1}^{k-1}
\alpha_i t_i U_i + \sum_{i=1}^{j-1} \beta_i t_i U_i \leq t_j$ in
Definition~\ref{def:kpoints-k2u}.  However, since the tests are
different, they are used for different situations and the resulting
bounds and properties are also different.

In Definition~\ref{def:kpoints-k2u}, the $k$-point effective
schedulability test is a sufficient schedulability test by testing
only $k$ time points, defined by the $k-1$ higher-priority tasks and
task $\tau_k$.  These $k-1$ points defined by the $k-1$
higher-priority tasks can be arbitrary as long as the corresponding
$\alpha_i > 0$ and $\beta_i > 0$ can be defined. In
Definition~\ref{def:kpoints}, the $k-1$ points defined by the $k-1$
higher-priority tasks have to be the last release times of the
highest-priority tasks, and the $k-1$ higher-priority tasks have to be
indexed according to their last release time before $t_k$.  In
Definition~\ref{def:kpoints}, the last release time ordering $\pi$ is
assumed to be given. In some cases, this ordering can be easily
obtained. However, in some of the cases in our demonstrated task
models in \cite{DBLP:journals/corr/abs-kRTA}, the last release ordering cannot be 
defined. It may seem that 
we have to test all possible last release time orderings and
take the worst case. Fortunately, finding the
worst-case ordering is not a difficult problem, which requires to sort
the $k-1$ higher-priority tasks under a simple criteria. Therefore,
before adopting the \frameworkkq{} framework, we have to know whether we
can obtain the last release time ordering or we have to consider a
pessimistic ordering for the higher priority tasks.

The frameworks assume that the corresponding coefficients $\alpha_i$
and $\beta_i$ in Definitions~\ref{def:kpoints-k2u},~\ref{def:kpoints},
and~\ref{def:kpoints-response} are given. How to derive them depends
on the task models and the scheduling policies.  Provided that these
coefficients $\alpha_i$, $\beta_i$, $C_i$, $U_i$ for every higher
priority task $\tau_i \in hp(\tau_k)$ are given, we can
find the worst-case assignments of the values $t_i$ for
the higher-priority tasks $\tau_i \in hp(\tau_k)$. Therefore, in case
Definition~\ref{def:kpoints-k2u} is adopted, changing $t_i$ affects the
values $\alpha_i t_i U_i$ and $\beta_i t_i U_i$; in case
Definitions~\ref{def:kpoints} and~\ref{def:kpoints-response} are
adopted, changing $t_i$ only affects the value $\alpha_i t_i U_i$. 
By using the above approach, we can analyze (1)
the response time by finding the extreme case for a given $C_k$ (under
Definition~\ref{def:kpoints-response}), or (2) the schedulability by
finding the extreme case for a given $C_k$ and $t_k$  (under
Definitions~\ref{def:kpoints-k2u} and Definition~\ref{def:kpoints}). 

In Section~\ref{sec:use-framework}, we will give a comparison about
the difference of Definition~\ref{def:kpoints-k2u} and Definition~\ref{def:kpoints} based on uniprocessor schedulability tests
for sporadic tasks.

\subsection{Properties of \frameworkku{}}

By using the property defined in Definition~\ref{def:kpoints-k2u}, we
can have the following lemmas in the \frameworkku{} framework
\cite{DBLP:journals/corr/abs-1501.07084,DBLP:conf/rtss/ChenHL15}. All the proofs of the
following lemmas are in
\cite{DBLP:journals/corr/abs-1501.07084,DBLP:conf/rtss/ChenHL15}.

\begin{lemma}
\label{lemma:framework-constrained-k2u}
For a given $k$-point effective schedulability test of a scheduling 
algorithm, defined in
Definition~\ref{def:kpoints-k2u},
in which $0 < t_k$ and $0 < \alpha_i \leq \alpha$, and $0 < \beta_i \leq \beta$ for any
$i=1,2,\ldots,k-1$, task $\tau_k$ is schedulable by the scheduling
algorithm if the following condition holds 
\begin{equation}
\label{eq:schedulability-constrained-k2u}
\frac{C_k}{t_k} \leq \frac{\frac{\alpha}{\beta}+1}{\prod_{j=1}^{k-1} (\beta U_j + 1)} - \frac{\alpha}{\beta}.
\end{equation}
\end{lemma}

\begin{lemma}
\label{lemma:framework-totalU-constrained-k2u}
For a given $k$-point effective schedulability test of a scheduling
algorithm, defined in
Definition~\ref{def:kpoints-k2u},
in which $0 < t_k$ and $0 < \alpha_i \leq \alpha$ and $0 < \beta_i \leq \beta$ for any
$i=1,2,\ldots,k-1$, task $\tau_k$ is schedulable by the scheduling
algorithm if 
\begin{equation}
\label{eq:schedulability-totalU-constrained-k2u}
\frac{C_k}{t_k} + \sum_{i=1}^{k-1}U_i \leq \frac{(k-1)((\alpha+\beta)^{\frac{1}{k}}-1)+((\alpha+\beta)^{\frac{1}{k}}-\alpha)}{\beta}.
\end{equation}
\end{lemma}

\begin{lemma}
\label{lemma:framework-totalU-exclusive-k2u}
For a given $k$-point effective schedulability test of a scheduling
algorithm, defined in
Definition~\ref{def:kpoints-k2u},
in which $0 < t_k$ and $0 < \alpha_i \leq \alpha$ and $0 < \beta_i \leq \beta$ for any
$i=1,2,\ldots,k-1$, task $\tau_k$ is schedulable by the scheduling
algorithm if 
\begin{equation}
\label{eq:schedulability-totalU-exclusive-k2u}
\beta \sum_{i=1}^{k-1}U_i \leq \ln(\frac{\frac{\alpha}{\beta}+1}{\frac{C_k}{t_k}+\frac{\alpha}{\beta}}).
\end{equation}
\end{lemma}

\begin{lemma}
\label{lemma:framework-general-k2u}
For a given $k$-point effective schedulability test of a fixed-priority scheduling
algorithm, defined in
Definition~\ref{def:kpoints-k2u}, task $\tau_k$ is schedulable by the scheduling algorithm,
in which $0 < t_k$ and $0 < \alpha_i$ and $0 < \beta_i$ for any
$i=1,2,\ldots,k-1$, 
 if
the following condition holds 
\begin{equation}
\label{eq:schedulability-general-k2u}
0 < \frac{C_k}{t_k} \leq 1 -  \sum_{i=1}^{k-1}  \frac{U_i(\alpha_i
  +\beta_i)}{\prod_{j=i}^{k-1} (\beta_jU_j + 1)}.
\end{equation}
\end{lemma}

\subsection{Properties of \frameworkkq{}}
By using the property defined in Definition~\ref{def:kpoints}, we
can have the following lemmas in the \frameworkkq{} framework
\cite{DBLP:journals/corr/abs-kRTA,DBLP:conf/rtss/ChenHL16}.
All the proofs of the
following lemmas are in
\cite{DBLP:journals/corr/abs-kRTA,DBLP:conf/rtss/ChenHL16}.

\begin{lemma}
\label{lemma:framework-general-schedulability}
For a given $k$-point last-release schedulability test, defined in
Definition~\ref{def:kpoints}, of a scheduling 
algorithm,
in which $0 < \alpha_i$, and $0 < \beta_i$ for any
$i=1,2,\ldots,k-1$, $0 < t_k$, $\sum_{i=1}^{k-1}\alpha_i U_i \leq 1$, and $\sum_{i=1}^{k-1}
\beta_i C_i \leq t_k$, task $\tau_k$ is schedulable by the
fixed-priority scheduling
algorithm if the following condition holds
\begin{equation}
\label{eq:schedulability-general}
\frac{C_k}{t_k} \leq 1 - \sum_{i=1}^{k-1}\alpha_i U_i - \frac{\sum_{i=1}^{k-1} (\beta_i C_i - \alpha_i U_i (\sum_{\ell=i}^{k-1}  \beta_\ell C_\ell) )}{t_k}.
\end{equation}
\end{lemma}

It may seem at first glance that we need to test all the possible
orderings. Fortunately, with the following lemma, we can safely
consider only one specific ordering of the $k-1$ higher priority
tasks.
\begin{lemma}
  \label{lemma:general-sorting}
  The worst-case ordering $\pi$ of the $k-1$ higher-priority tasks under the schedulability condition in Eq.~\eqref{eq:schedulability-general} in Lemma~\ref{lemma:framework-general-schedulability} is to order the tasks in a non-increasing order of $\frac{\beta_i C_i}{\alpha_i U_i}$,
 in which $0 < \alpha_i$ and $0 < \beta_i$ for any $i=1,2,\ldots,k-1$, and $0 < t_k$.
\end{lemma}

The analysis in Lemma~\ref{lemma:framework-general-schedulability} uses the execution time and the utilization of the tasks in $hp(\tau_k)$ to build an upper bound of $C_k/t_k$ for  schedulability tests. It is also very convenient in real-time systems to build schedulability tests only based on utilization of the tasks. We explain how to achieve that in the following lemmas under the assumptions that $0 < \alpha_i \leq \alpha$, and $0 < \beta_i C_i \leq \beta U_i t_k$ for any $i=1,2,\ldots,k-1$. 
These lemmas are useful when we are interested to derive utilization bounds, speed-up factors, resource augmentation factors, etc., for a given scheduling policy by defining the coefficients $\alpha$ and $\beta$ according to the scheduling policies independently from the detailed parameters of the tasks. 
 Since the property repeats in all the statements, we make a formal definition before presenting the lemmas.
\begin{definition}
  \label{def:alpha-upper-bound}
  Lemmas~\ref{lemma:framework-constrained-schedulability} to
  \ref{lemma:framework-totalU-constrained} are based on the
  following $k$-point last-release schedulability test of a scheduling
  algorithm, defined in Definition~\ref{def:kpoints}, in which $0 <
  \alpha_i \leq \alpha$, and $0 < \beta_i C_i \leq \beta U_i t_k$ for
  any $i=1,2,\ldots,k-1$, $0 < t_k$, $\alpha\sum_{i=1}^{k-1}U_i \leq
  1$, and $\beta\sum_{i=1}^{k-1}U_i \leq 1$.
\end{definition}

\begin{lemma}
\label{lemma:framework-constrained-schedulability}
For a given $k$-point last-release schedulability test of a scheduling 
algorithm, with the properties in Definition~\ref{def:alpha-upper-bound}, 
task $\tau_k$ is schedulable by the scheduling
algorithm if the following condition holds
{\small \begin{align}
\label{eq:schedulability-constrained}
\frac{C_k}{t_k} \leq &1 - (\alpha+\beta)\sum_{i=1}^{k-1} U_i + \alpha\beta\sum_{i=1}^{k-1} U_i (\sum_{\ell=i}^{k-1}  U_\ell)\\
= &1 - (\alpha+\beta)\sum_{i=1}^{k-1} U_i + 0.5\alpha\beta\left((\sum_{i=1}^{k-1} U_i)^2 +(\sum_{i=1}^{k-1} U_i^2)\right)
\label{eq:schedulability-constrained-2}
\end{align}}
\end{lemma}

\begin{figure*}[t]
	\begin{center}
      \begin{tikzpicture}[scale=0.9,transform shape]
        \path \practica {1}{\underline{\bf Demonstrated Applications:}
          \begin{tabular}{ll}
          Sec. 5:& Uniprocessor sporadic tasks\\
          Sec. 6:& Multiprocessor global RM\\
          \multicolumn{2}{l}{and several others in \cite{DBLP:journals/corr/abs-1501.07084,DBLP:conf/rtss/ChenHL15,DBLP:journals/corr/abs-kRTA,DBLP:conf/rtss/ChenHL16}. }
          \end{tabular}
        };
        \path (p1.west)+(2.2,-7.8) node(p2)[materia, rounded
        rectangle,text width=7.2em]{{\bf $U_i, \forall i < k$\\ $C_i,
            \forall i < k$\\$\alpha_i, \forall i<k$\\$\beta_i, \forall
            i <k$\\$C_k$\\ $t_k$ {\scriptsize(for Lemmas \ref{lemma:framework-general-schedulability}-\ref{lemma:framework-totalU-constrained})}}}; 
        \path (p2.north)+(1.5,0.45) node[text width=16em]{Derive
          parameters\\by \underline{Definitions 3 or 4}};
        \path (p2.east)+(3,0) node(p3)[practica,fill=blue!40,text width=5em,text centered]{\large{\bf $k^2Q$\\ framework}};

        \path (p3.east)+(4.5,+3) node(p4)[practica,fill=green!45,minimum width=6em,text width=5em,text centered]{Quadratic bound}; 
        \path (p4.south)+(0,-1) node(p5)[practica,fill=green!45,minimum width=6em,text width=5em,text centered]{Other utilization bounds}; 
        \path (p5.south)+(0,-1)
        node(p6)[practica,fill=green!45,minimum width=6em,text
        width=5em,text centered]{Response-time test}; 
        
        \path [line, ->] (p1.west)+(0,0.8) -- +(-1,0.8) node[black, rotate=90,
        yshift=0.3cm, xshift=-4cm]{\footnotesize{define the
            last release time ordering $\pi$ or use Lemma~\ref{lemma:general-sorting} or \ref{lemma:general-response-sorting}}} -- + (-1, -7.8) -- (p2.west);
        \path [line, ->] (p2.east) -- (p3.west);
        \path [line, ->] (p3.east)+(0,0.3) -- node[rotate=45,yshift=0.3cm,black]{Lemma \ref{lemma:framework-general-schedulability}} (p4.west);
        \path [line, ->] (p3.east)+(0,0) -- node[rotate=25,yshift=0.3cm,black]{Lemmas \ref{lemma:framework-constrained-schedulability}-~\ref{lemma:framework-totalU-constrained}} (p5.west);
        \path [line, ->] (p3.east)+(0,-0.3) --
        node[rotate=3,yshift=0.3cm,black]{Lemma \ref{lemma:framework-general-response}} (p6.west);
        \path (p1.west)+(2.8,-3) node(p12)[materia, rounded rectangle,text width=6em]{{\bf $U_i, \forall i < k$\\$\alpha_i, \forall i<k$\\$\beta_i, \forall i <k$\\$C_k, t_k$}}; 
        \path (p12.north)+(1.5,0.45) node[text width=16em]{Derive parameters\\by \underline{Definition 1}};
        \path (p12.east)+(3,0) node(p13)[practica,fill=blue!10,text width=5em,text centered]{\large{\bf \frameworkku{}\\ framework}};

        \path (p13.east)+(4.5,+3) node(p14)[practica,fill=green!15,minimum width=6em,text width=5em,text centered]{Hyperbolic bound}; 
        \path (p14.south)+(0,-1) node(p15)[practica,fill=green!15,minimum width=6em,text width=5em,text centered]{Other utilization bounds}; 
        \path (p15.south)+(0,-1) node(p16)[practica,fill=green!15,minimum width=6em,text width=5em,text centered]{Extreme points test}; 
        
        \path [line, ->] (p1.west) -- +(-0.3,0) node[black, rotate=90, yshift=0.3cm, xshift=-1.6cm]{\footnotesize{define $t_i, \forall i < k$ and order $k-1$ tasks}} -- + (-0.3, -3) -- (p12.west);
        \path [line, ->] (p12.east) -- (p13.west);
        \path [line, ->] (p13.east)+(0,0.3) -- node[rotate=40,yshift=0.3cm,black]{Lemma \ref{lemma:framework-constrained-k2u}} (p14.west);
        \path [line, ->] (p13.east)+(0,0) -- node[rotate=25,yshift=0.3cm,black]{Lemmas \ref{lemma:framework-totalU-constrained-k2u}\&~\ref{lemma:framework-totalU-exclusive-k2u}} (p15.west);
        \path [line, ->] (p13.east)+(0,-0.3) -- node[rotate=5,yshift=0.3cm,black]{Lemma \ref{lemma:framework-general-k2u}} (p16.west);
      \end{tikzpicture}    
	\end{center}
\vspace{-2mm}
\caption{The \frameworkku{} and \frameworkkq{} frameworks. }
\label{fig:framework}
\end{figure*}
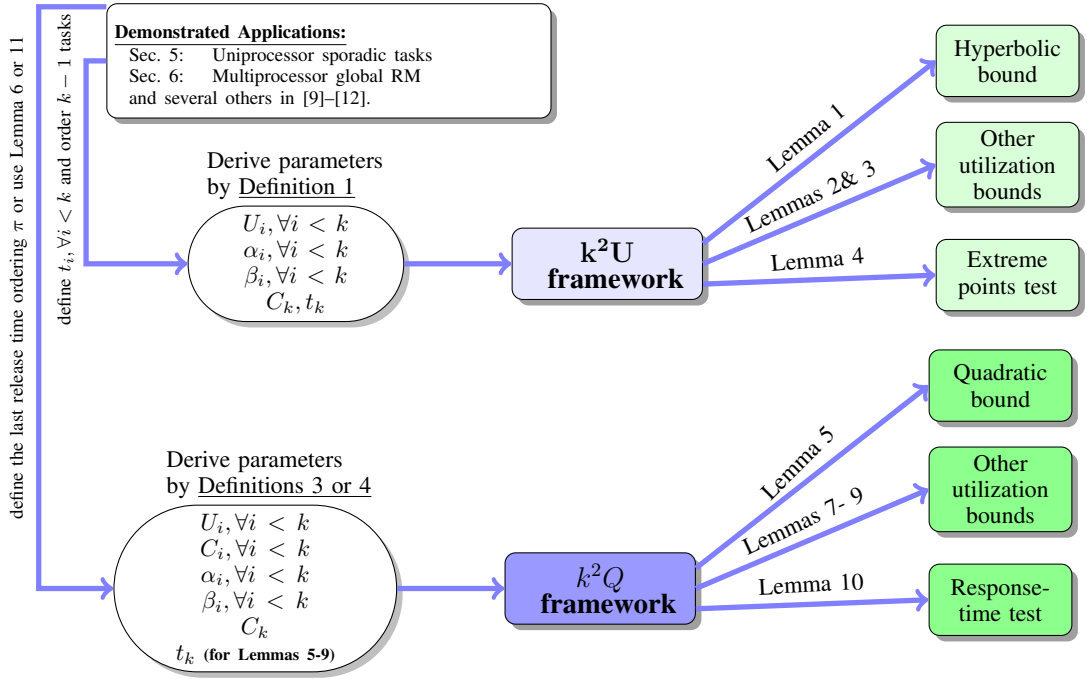

\begin{lemma}
\label{lemma:framework-totalU-exclusive}
For a given $k$-point last-release schedulability test of a scheduling 
algorithm, with the properties in Definition~\ref{def:alpha-upper-bound}, 
task $\tau_k$ is schedulable by the scheduling
algorithm if 
\begin{equation}\small
\label{eq:schedulability-totalU-exclusive}
\sum_{i=1}^{k-1}U_i \leq \left(\frac{k-1}{k}\right)\left( \dfrac{\alpha+\beta-\sqrt{(\alpha+\beta)^2-2 \alpha\beta (1-\frac{C_k}{t_k})\frac{k}{k-1}}}{\alpha\beta}\right).
\end{equation}
\end{lemma}

\begin{lemma}
\label{lemma:framework-totalU-constrained}
For a given $k$-point last-release schedulability test of a scheduling 
algorithm, with the properties in Definition~\ref{def:alpha-upper-bound}, 
provided that $\alpha+\beta \geq 1$, 
then task $\tau_k$ is schedulable by the scheduling
algorithm if  
{\small
\begin{align}
&\frac{C_k}{t_k} + \sum_{i=1}^{k-1} U_i \leq \nonumber\\
&\begin{cases}\label{eq:schedulability-totalU-constrained}
  \left(\frac{k-1}{k}\right)\left( \dfrac{\alpha+\beta-\sqrt{(\alpha+\beta)^2-2 \alpha\beta \frac{k}{k-1}}}{\alpha\beta}\right), &
  \begin{array}{l}
    \mbox{ if  } k > \frac{(\alpha+\beta)^2-1}{\alpha^2+\beta^2-1} \\    
    \mbox{ and  }\alpha^2+\beta^2 > 1
  \end{array}\\
1 + \frac{(k-1)((\alpha+\beta-1) - \frac{1}{2}(\alpha+\beta)^2+0.5) }{k\alpha\beta} & \mbox{ otherwise}
\end{cases}
 \end{align} 
}
\end{lemma}

The right-hand side of
Eq.~\eqref{eq:schedulability-totalU-constrained} (when
$\alpha^2+\beta^2 > 1$) decreases with
respect to $k$. Similarly, the right-hand side of
Eq.~\eqref{eq:schedulability-totalU-exclusive} also decreases with
respect to $k$. Therefore, for evaluating the utilization bounds, it
is alway safe to take $k\rightarrow \infty$ as a safe upper bound. The
right-hand side of Eq.~\eqref{eq:schedulability-totalU-exclusive}
converges to $\frac{\alpha+\beta-\sqrt{\alpha^2+\beta^2+2
    \alpha\beta\frac{C_k}{t_k}}}{\alpha\beta}$ when $k\rightarrow
\infty$.  The right-hand side of
Eq.~\eqref{eq:schedulability-totalU-constrained}  (when
$\alpha^2+\beta^2 > 1$) converges to
$\frac{\alpha+\beta-\sqrt{\alpha^2+\beta^2}}{\alpha\beta}$ when
$k\rightarrow \infty$.


The following two lemmas are from the $k$-point last-release response time analysis, defined in
Definition~\ref{def:kpoints-response}.
\begin{lemma}
\label{lemma:framework-general-response}
For a given $k$-point last-release response time analysis of a scheduling 
algorithm, defined in
Definition~\ref{def:kpoints-response},
in which $0 < \alpha_i \leq \alpha$, $0 < \beta_i \leq \beta$ for any
$i=1,2,\ldots,k-1$, $0 < t_k$ and $\sum_{i=1}^{k-1}
\alpha_i U_i < 1$, the response time to execute $C_k$ for task
$\tau_k$ is at most
\begin{equation}
\label{eq:schedulability-general-response}
\frac{C_k+ \sum_{i=1}^{k-1} \beta_i C_i - \sum_{i=1}^{k-1} \alpha_i U_i (\sum_{\ell=i}^{k-1} \beta_{\ell} C_{\ell})}{1-\sum_{i=1}^{k-1} \alpha_i U_i}.
\end{equation}
\end{lemma}

\begin{lemma}
  \label{lemma:general-response-sorting}
  The worst-case ordering $\pi$ of the $k-1$ higher-priority tasks
  under the response bound in
  Eq.~\eqref{eq:schedulability-general-response} in Lemma~\ref{lemma:framework-general-response} is to order the tasks in a non-increasing order of $\frac{\beta_i C_i}{\alpha_i U_i}$,
 in which $0 < \alpha_i$ and $0 < \beta_i$ for any $i=1,2,\ldots,k-1$, $0 < t_k$.
\end{lemma}

\section{How to Use the Frameworks}
\label{sec:use-framework}

The \frameworkku{} and \frameworkkq{} frameworks can be used by a wide range of applications, as
long as the users can properly specify the corresponding task
properties $C_i$ (in case of \frameworkkq{}) and $U_i$ and the constant
coefficients $\alpha_i$ and $\beta_i$ of every higher priority task
$\tau_i$. The choice of the parameters $\alpha_i$ and $\beta_i$
affects the quality of the resulting schedulability bounds.  However,
deriving the \emph{good} settings of $\alpha_i$ and $\beta_i$ is
not the focus of the frameworks. The frameworks do not care how
the parameters $\alpha_i$ and $\beta_i$ are obtained. It simply
derives the bounds according to the given parameters $\alpha_i$ and
$\beta_i$, regardless of the settings of $\alpha_i$ and $\beta_i$. The
correctness of the settings of $\alpha_i$ and $\beta_i$ is not
verified by the frameworks. Figure \ref{fig:framework} provides an overview of the procedures.

The ignorance of the settings of $\alpha_i$ and $\beta_i$ actually
leads to the elegance and the generality of the frameworks, which work
as long as Definitions~\ref{def:kpoints-k2u}, \ref{def:kpoints}, or
\ref{def:kpoints-response} can be successfully constructed for the
sufficient schedulability test or the response time analysis.  The
other approaches in
\cite{journals/tc/LeeSP04,DBLP:dblp_journals/tc/BurchardLOS95,HanTyan-RTSS97}
also have similar observations by testing only several time points in
the TDA schedulability analysis based on
Eq.~\eqref{eq:exact-test-constrained-deadline} in their problem
formulations. However, as these approaches in
\cite{journals/tc/LeeSP04,DBLP:dblp_journals/tc/BurchardLOS95,HanTyan-RTSS97}
seek for the total utilization bounds, they have limited applications
and are less flexible. For example, they are typically not applicable
directly when considering sporadic real-time tasks with arbitrary
deadlines or multiprocessor systems.

The \frameworkku{} and \frameworkkq{} frameworks provide comprehensive means for the users to
almost automatically convert the pseudo polynomial-time tests (or even
exponential-time tests) into polynomial-time tests with closed
mathematical forms.  With the availability of the \frameworkku{} and \frameworkkq{} frameworks,
the hyperbolic bounds, quadratic bounds, or speedup factors can be
almost automatically derived by adopting the provided lemmas in Section~\ref{sec:framework}
as long as the safe upper bounds $\alpha$ and $\beta$ to cover all the
possible settings of $\alpha_i$ and $\beta_i$ for the schedulability
test or the response-time analysis can be derived, regardless of the
task model or the platforms.

The above characteristics and advantages over other approaches have
been already discussed in
\cite{DBLP:journals/corr/abs-1501.07084,DBLP:conf/rtss/ChenHL15,DBLP:journals/corr/abs-kRTA,DBLP:conf/rtss/ChenHL16}. However,
between these two frameworks, it is unclear
whether both are needed or only one of them is important.

As the simplest example, consider the test of task $\tau_2$ with
$T_2=1$ in an implicit-deadline sporadic task set in uniprocessor RM scheduling. Suppose that task
$\tau_1$ has utilization $U_1=0.3$. If we only use the utilization of
the higher-priority tasks as the means of testing, modeling the
schedulability test in Definition~\ref{def:kpoints} is less precise
since we may have to inflate and set $C_1$ properly according to the
given priority assignment. Using Definition~\ref{def:kpoints-k2u} with
$t_1=0.7$ leads to $C_1=0.21$, but using Definition~\ref{def:kpoints}
with any $0 < t_1 \leq 1$ can only be feasible if we set $C_1$ to
$0.3$. Therefore, for such cases, we can only be safe by putting $C_i
= t_k U_i$, and, therefore, using \frameworkkq{} is more pessimistic
than using \frameworkku{}.

In the above example, it may seem at first glance that the test in the
\frameworkku{} framework is better than the test in the \frameworkkq{}
framework. However, this observation can only hold if a
schedulability test can be applicable to satisfy
Definition~\ref{def:kpoints-k2u} and Definition~\ref{def:kpoints}.

\begin{figure}[t]
   \centering
  \includegraphics[width=\columnwidth]{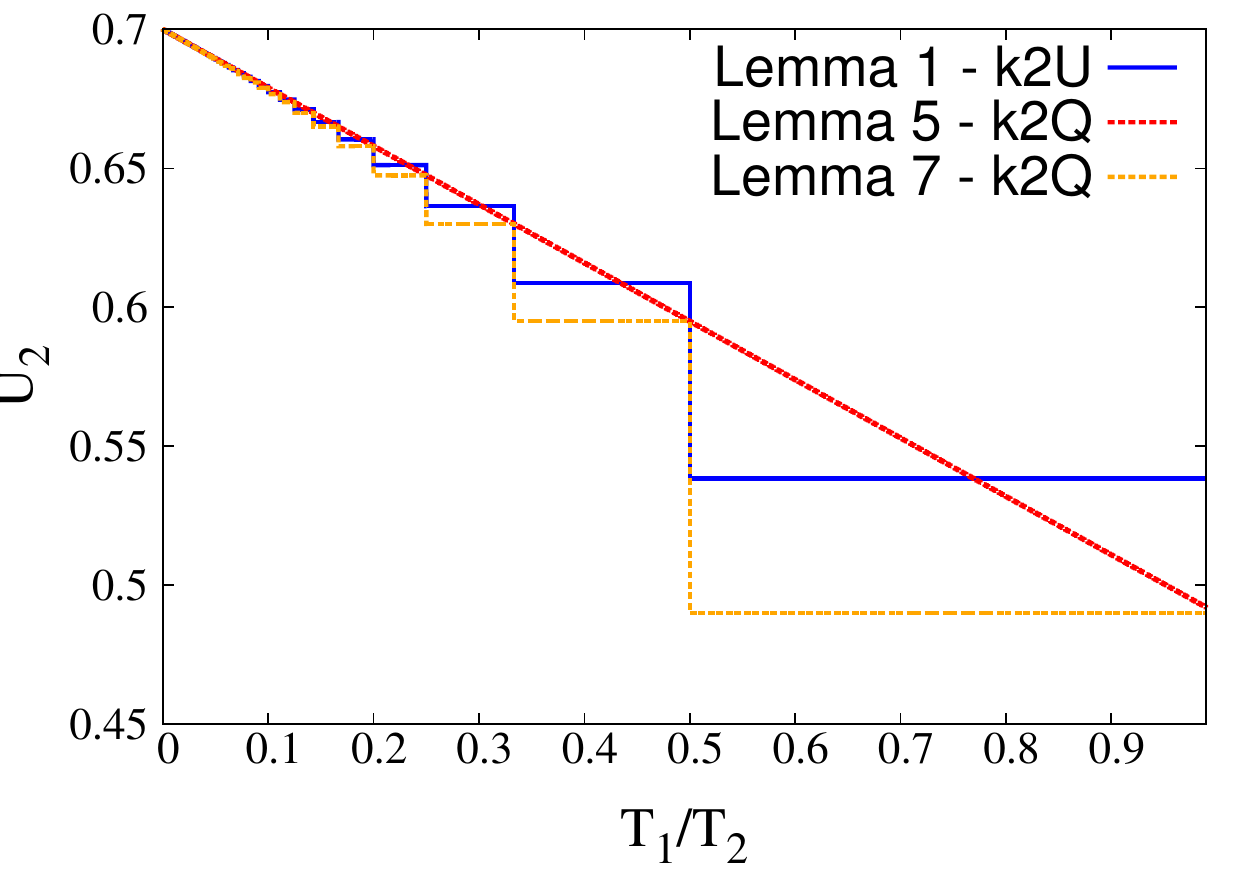}
  \caption{Adopting different tests from \frameworkku{} and \frameworkkq{} for RM uniprocessor scheduling with $k=2$.}
    \label{fig:2tasks-example-rm}
\end{figure}

We test the above case with different settings of $\frac{T_1}{T_2}$
with $T_2 > T_1$ when $U_1$ is
$0.3$. Figure~\ref{fig:2tasks-example-rm} illustrates the maximum
utilization of task $\tau_2$ by using different tests from the two
frameworks. In such a case, we can clearly define $t_1$ as
$\ceiling{\frac{T_2}{T_1}-1} T_1$. Therefore, $\alpha_1$ is $1$
and $\beta_1$ is set to $\frac{1}{\ceiling{\frac{T_2}{T_1}-1}}$ when
adopting Lemma \ref{lemma:framework-constrained-k2u} from
\frameworkku{}. Moreover, $\alpha_1$ is $1$ and $\beta_1$ is set to
$\frac{1}{\ceiling{\frac{T_2}{T_1}-1}}$ when adopting Lemma
\ref{lemma:framework-constrained-schedulability} from \frameworkkq{}.

As shown in Figure~\ref{fig:2tasks-example-rm}, when we adopt only
utilizations of the higher-priority task, i.e., Lemma
\ref{lemma:framework-constrained-k2u} from \frameworkku{} and Lemma
\ref{lemma:framework-constrained-schedulability} from \frameworkkq{},
the results from \frameworkku{} are always better. However, the results
of Lemma \ref{lemma:framework-constrained-k2u} from \frameworkku{} and
Lemma \ref{lemma:framework-general-schedulability} from \frameworkkq{}
are not comparable. 

Therefore, there is no clear dominator between these two frameworks.
Moreover, there are also cases, in which formulating the test by using
the \frameworkku{} framework is not possible (c.f. the results in
Theorems~\ref{theorem:response-time-sporadic} and
\ref{thm:multiprocessor-grm-sporadic-tight}). These cases may even
start from schedulability tests with exponential-time complexity. We
have successfully demonstrated three examples in
\cite{DBLP:journals/corr/abs-kRTA} by using the \frameworkkq{} framework
to derive polynomial-time tests with approximation guarantees. In
those demonstrated cases, either the \frameworkku{} framework cannot
be applied or with worse results (since different exponential-time or
pseudo-polynomial-time schedulability tests are applied).

\section{Analysis for Sporadic Task Models}
\label{sec:sporadic}

This section examines the applicability of the \frameworkku{} and
\frameworkkq{} frameworks to derive utilization-based schedulability
analysis and response-time analysis for sporadic task systems in
uniprocessor systems. We will consider constrained-deadline systems in
Section~\ref{sec:constrained-deadline} and arbitrary-deadline systems
in Section~\ref{sec:arbitrary-deadline}.  For a specified
fixed-priority scheduling algorithm, let $hp(\tau_k)$ be the set of
tasks with higher priority than $\tau_k$. We now classify the task set
$hp(\tau_k)$ into two subsets:
\begin{itemize}
\item $hp_1(\tau_k)$ consists of the higher-priority tasks with periods
  smaller than $D_k$.
\item $hp_2(\tau_k)$ consists of the higher-priority tasks with periods
  larger than or equal to $D_k$.
\end{itemize}
For the rest of this section, we will implicitly assume $C_k > 0$.

\subsection{Constrained-Deadline}
\label{sec:constrained-deadline}

For a constrained-deadline task $\tau_k$, the schedulability test in
Eq.~\eqref{eq:exact-test-constrained-deadline}  is equivalent to the
verification of the existence of $0 < t \leq D_k$ such that
\begin{equation}
  \label{eq:exact-test-constrained-deadline-2}
 C_k + \sum_{\tau_i \in hp_2(\tau_k)} C_i + \sum_{\tau_i \in hp_1(\tau_k)} \ceiling{\frac{t}{T_i}}C_i \leq t.  
\end{equation}
We can then create a virtual sporadic task $\tau_k'$ with execution
time $C_k'=C_k + \sum_{\tau_i \in hp_2(\tau_k)} C_i$, relative
deadline $D_k'=D_k$, and period $T_k'=D_k$. It is
clear that the schedulability test to verify the schedulability of
task $\tau_k'$ under the interference of the higher-priority tasks
$hp_1(\tau_k)$ is the same as that of task $\tau_k$ under the
interference of the higher-priority tasks $hp(\tau_k)$.
For notational brevity, suppose that there are $k^*-1$ tasks, indexed
as $1,2,\ldots,k^*-1$, in
$hp_1(\tau_k)$.

\noindent\textbf{Adopting \frameworkku{}}: \hspace{1cm}
Setting $t_i =\left (\ceiling{\frac{D_k}{T_i}}-1\right)T_i$ for every
task $\tau_i$ in $hp_1(\tau_k)$, and indexing the
tasks in a non-decreasing order of $t_i$ lead to the satisfaction of
Definition~\ref{def:kpoints-k2u} with $\alpha_i=1$ and $0 < \beta_i
\leq 1$.
Therefore, we can apply Lemmas~\ref{lemma:framework-constrained-k2u}
and~\ref{lemma:framework-totalU-constrained-k2u} to obtain the
following theorem. 
 
\begin{theorem}
\label{theorem:sporadic-general-k2u}
Task $\tau_k$ in a sporadic task system with constrained deadlines is
schedulable by the fixed-priority scheduling algorithm if
\begin{equation}
\label{eq:schedulability-sporadic-k2u-any-a}
(\frac{C_k'}{D_k}+1) \prod_{\tau_j \in hp_1(\tau_k)} (U_j + 1)\leq 2
\end{equation}
or
\begin{equation}
\label{eq:schedulability-sporadic-k2u-any-b}
\frac{C_k'}{D_k}+  \sum_{\tau_j \in hp_1(\tau_k)}U_j\leq  k^*(2^{\frac{1}{k^*}}-1).
\end{equation}
\end{theorem}

\begin{corollary}
\label{corollary-rm-k2u}
Task $\tau_k$ in a sporadic task system with implicit deadlines is
schedulable by the RM scheduling algorithm if
Lemmas \ref{lemma:framework-totalU-constrained-k2u}
and \ref{lemma:framework-totalU-exclusive-k2u}
holds by setting
$\frac{C_k}{t_k}$ as $U_k$, $\alpha=1$, and $\beta=1$.
\end{corollary}

The above result in Corollary~\ref{corollary-rm-k2u} leads to the
utilization bound $\ln{2}$ (by using
Lemma~\ref{lemma:framework-totalU-constrained-k2u} with $\alpha=1$ and
$\beta=1$) for RM scheduling, which is the same as the Liu and Layland
bound $\ln{2}$ \cite{liu1973scheduling}. It also leads to the
hyperbolic bound for RM scheduling by Bini and Buttazzo
\cite{bini2003rate} when adopting
Theorem~\ref{theorem:sporadic-general-k2u} directly.

\noindent\textbf{Adopting \frameworkkq{}}: \hspace{1cm} 
Setting $t_i =\left (\ceiling{\frac{D_k}{T_i}}-1\right)T_i$ for every
task $\tau_i$ in $hp_1(\tau_k)$, and indexing the
tasks in a non-decreasing order of $t_i$ leads to the satisfaction of
Definition~\ref{def:kpoints} with $\alpha_i=1$ and
$\beta_i=1$.  For such a case, the last release ordering is well-defined by the sorting of the tasks above.
Therefore, we can use
Lemma~\ref{lemma:framework-general-schedulability} to derive the
following theorem.

\begin{theorem}
\label{theorem:sporadic-general}
Task $\tau_k$ in a sporadic task system with constrained deadlines is
schedulable by the fixed-priority scheduling algorithm if
$\sum_{i=1}^{k^*-1}\frac{C_i}{D_k} \leq 1$ and
\begin{equation}
\label{eq:schedulability-sporadic-any-a}
\frac{C_k'}{D_k} \leq 1-\sum_{i=1}^{k^*-1}U_i-\sum_{i=1}^{k^*-1}\frac{C_i}{D_k}+\frac{\sum_{i=1}^{k^*-1} U_i (\sum_{\ell=i}^{k^*-1}  C_\ell)}{D_k},
\end{equation}
in which the $k^*-1$ higher priority tasks in $hp_1(\tau_k)$ are indexed
in a non-decreasing order of $\left(\ceiling{\frac{D_k}{T_i}}-1\right)T_i$.
\end{theorem}

\begin{corollary}
\label{corollary-rm}
Task $\tau_k$ in a sporadic task system with implicit deadlines is
schedulable by the RM scheduling algorithm if
Lemmas~\ref{lemma:framework-general-schedulability},~\ref{lemma:framework-constrained-schedulability},
\ref{lemma:framework-totalU-exclusive},
or~\ref{lemma:framework-totalU-constrained} holds by setting
$\frac{C_k}{t_k}$ as $U_k$, $\alpha=1$, and $\beta=1$.
\end{corollary}

The above result in Corollary~\ref{corollary-rm} leads to the
utilization bound $2-\sqrt{2}$ (by using
Lemma~\ref{lemma:framework-totalU-constrained} with $\alpha=1$ and
$\beta=1$) for RM scheduling, which is worse than the existing
Liu and Layland bound $\ln{2}$ \cite{liu1973scheduling}. 

Moreover, the above utilization bound $2-\sqrt{2}$ has been also
provided by Abdelzaher et al. \cite{DBLP:journals/tc/AbdelzaherSL04}
for uniprocessor deadline-monotonic scheduling when an aperiodic task
may generate different task instances (jobs) with different
combinations of execution times and minimum inter-arrival times. Such
a model is a more general model than the sporadic task model. Under
such a setting, the \frameworkku{} framework cannot be used, whereas
the \frameworkkq{} framework is very suitable.

\subsection{Arbitrary-Deadline}
\label{sec:arbitrary-deadline}

The schedulability analysis for arbitrary-deadline sporadic task sets
is to use a \emph{busy-window} concept to evaluate the worst-case
response time \cite{DBLP:conf/rtss/Lehoczky90}. That is, we release
all the higher-priority tasks together with task $\tau_k$ at time $0$
and all the subsequent jobs are released as early as possible by
respecting to the minimum inter-arrival time. The busy window finishes
when a job of task $\tau_k$ finishes before the next release of a job
of task $\tau_k$. It has been shown in
\cite{DBLP:conf/rtss/Lehoczky90} that the worst-case response time of
task $\tau_k$ can be found in one of the jobs of task $\tau_k$ in the
busy window.
For the $h$-th job of task $\tau_k$ in the busy window, let  the finishing
time $R_{k,h}$ is the minimum $t$ such that
\[ 
h C_k + \sum_{i=1}^{k-1} \ceiling{\frac{t}{T_i}}C_i \leq t, 
\] 
and, hence, its response time is $R_{k,h}-(h-1)T_k$. The busy window
of task $\tau_k$ finishes on the $h$-th job if $R_{k,h} \leq h
T_k$. 

A simpler sufficient schedulability test for a task $\tau_k$ is to
test whether the length of the busy window is within $D_k$.  If so,
all invocations of task $\tau_k$ released in the busy window can
finish before their relative deadline. Such an observation has also
been adopted in \cite{conf:/rtns09/Davis}. Therefore, a sufficient
test is to verify whether
\begin{equation}
  \label{eq:sufficient-test-arbitrary-deadline}
\exists t \mbox{ with } 0 < t \leq D_k {\;\; and \;\;} \ceiling{\frac{t}{T_k}}C_k +
\sum_{\tau_i \in hp(\tau_k)} \ceiling{\frac{t}{T_i}}C_i \leq t. 
\end{equation}
If the condition in Eq.~\eqref{eq:sufficient-test-arbitrary-deadline} holds,
it implies that the busy window (when considering task $\tau_k$) is no
more than $D_k$, and, hence, task $\tau_k$ has worst-case response
time no more than $D_k$.

Similarly, we can use $hp_1(\tau_k)$ and $hp_2(\tau_k)$, as in
Section~\ref{sec:constrained-deadline}, and, then create a virtual
sporadic task $\tau_k'$ with execution time $C_k' =
\ceiling{\frac{D_k}{T_k}}C_k + \sum_{\tau_i \in hp_2(\tau_k)} C_i$,
relative deadline $D_k'=D_k$, and period $T_k'=D_k$. 
For notational brevity, suppose that there are $k^*-1$ tasks, indexed
as $1,2,\ldots,k^*-1$, in
$hp_1(\tau_k)$.

\noindent\textbf{Adopting \frameworkku{}}: \hspace{1cm}
Setting $t_i =\left (\ceiling{\frac{D_k}{T_i}}-1\right)T_i$, and indexing the
tasks in a non-decreasing order of $t_i$ leads to the satisfaction of
Definition~\ref{def:kpoints-k2u} with $\alpha_i=1$ and $\beta_i \leq
1$. Therefore, we can apply Lemmas~\ref{lemma:framework-constrained-k2u}
and~\ref{lemma:framework-totalU-constrained-k2u} to obtain the
following theorem. 

\begin{theorem}
\label{theorem:sporadic-arbitrary-k2u}
Task $\tau_k$ in a sporadic task system with arbitrary deadlines is
schedulable by the fixed-priority scheduling algorithm if
\begin{equation}
\label{eq:schedulability-sporadic-arbitrary-k2u-any-a}
(\frac{C_k'}{D_k}+1) \prod_{\tau_j \in hp_1(\tau_k)} (U_j + 1)\leq 2
\end{equation}
or
\begin{equation}
\label{eq:schedulability-sporadic-arbitrary-k2u-any-b}
\frac{C_k'}{D_k}+  \sum_{\tau_j \in hp_1(\tau_k)}U_j\leq  k^*(2^{\frac{1}{k^*}}-1).
\end{equation}
\end{theorem}

\noindent\textbf{Adopting \frameworkkq{}}: \hspace{1cm}  If we use the
busy-window concept to analyze the schedulability of task $\tau_i$ by
using Eq.~\eqref{eq:sufficient-test-arbitrary-deadline}, we can reach
the following theorem directly by
Lemma~\ref{lemma:framework-general-schedulability}.
\begin{theorem}
\label{theorem:schedulability-sporadic-arbitrary}
Task $\tau_k$ in a sporadic task system is
schedulable by the fixed-priority scheduling algorithm if
$\sum_{i=1}^{k^*-1}\frac{C_i}{D_k} \leq 1$ and
\begin{equation}
\label{eq:schedulability-sporadic-any-arbitrary-a}
\frac{C_k'}{D_k} \leq 1-\sum_{i=1}^{k^*-1}U_i-\sum_{i=1}^{k^*-1}\frac{C_i}{D_k}+\frac{\sum_{i=1}^{k^*-1} U_i (\sum_{\ell=i}^{k^*-1}  C_\ell)}{D_k},
\end{equation}
in which $C_k' = \ceiling{\frac{D_k}{T_k}}C_k + \sum_{\tau_i \in
  hp_2(\tau_k)} C_i$, and the $k^*-1$ higher priority tasks in $hp_1(\tau_k)$ are indexed
in a non-decreasing order of $\left(\ceiling{\frac{D_k}{T_i}}-1\right)T_i$.
\end{theorem}

Analyzing the schedulability by using
Theorem~\ref{theorem:schedulability-sporadic-arbitrary} can be good if
$\frac{D_k}{T_k}$ is small. However, as the busy window may be
stretched when $\frac{D_k}{T_k}$ is large, it may be too pessimistic.  
Suppose that $t_j = \left(\ceiling{\frac{R_{k,h}}{T_j}}-1\right)T_j$
for a higher priority task $\tau_j$. We index the tasks such that the
last release ordering $\pi$ of the $k-1$ higher priority tasks is with
$t_j \leq t_{j+1}$ for $j=1,2,\ldots,k-2$. Therefore, we know that
$R_{k,h}$ is upper bounded by finding the maximum
  \begin{equation}
    \label{eq:sporadic-arbitrary-objective-k}
   t_k = hC_k + \sum_{i=1}^{k-1} t_i U_i + \sum_{i=1}^{k-1} C_i,
  \end{equation}
with $0 \leq t_1 \leq t_2 \leq \cdots \leq t_{k-1} \leq t_{k}$ and
  \begin{align}
    \label{eq:sporadic-arbitrary-objective-k2}
    hC_k + \sum_{i=1}^{k-1}  t_i U_i + \sum_{i=1}^{j-1}  C_i > t_j, & \forall j=1,2,\ldots,k-1.
  \end{align}
Therefore, the above derivation of $R_{k,h}$ satisfies
Definition~\ref{def:kpoints-response} with $\alpha_i=1$, and
$\beta_i=1$ for any higher priority task $\tau_i$. However, it should
be noted that the last release time ordering $\pi$ is actually unknown
since $R_{k,h}$ is unknown. Therefore, we have to apply
Lemma~\ref{lemma:general-response-sorting} for such cases to obtain
the worst-case ordering.

\begin{lemma}
  \label{lemma:finishing-time-sporadic-h}
  Suppose that $\sum_{i=1}^{k-1}  U_i \leq 1$. Then, for any
  $h \geq 1$ and $C_k > 0$, we have
  \begin{equation}
    \label{eq:R-k-h}
     R_{k,h} \leq \frac{hC_k+ \sum_{i=1}^{k-1}  C_i - \sum_{i=1}^{k-1} U_i (\sum_{\ell=i}^{k-1}  C_{\ell})}{1-\sum_{i=1}^{k-1}  U_i},
  \end{equation}
  where the $k-1$ higher-priority tasks are ordered
  in a non-increasing order of their periods.
\end{lemma}

The worst-case response time for such cases can be set to $h=1$, in
which the detailed proof is in \cite{DBLP:journals/corr/abs-kRTA,DBLP:conf/rtss/ChenHL16}.
\begin{theorem}
  \label{theorem:response-time-sporadic}
  Suppose that $\sum_{i=1}^{k} U_i \leq 1$. The worst-case
  response time of task $\tau_k$ is at most
  \begin{equation}
    \label{eq:R-k}
    R_{k} \leq \frac{C_k+ \sum_{i=1}^{k-1}  C_i - \sum_{i=1}^{k-1} U_i (\sum_{\ell=i}^{k-1}  C_{\ell})}{1-\sum_{i=1}^{k-1}  U_i},
  \end{equation}
  where the $k-1$ higher-priority tasks are ordered
  in a non-increasing order of their periods.
\end{theorem}

\begin{corollary}
  \label{corollary:arbitrary-response-schedulability}
  Task $\tau_k$ in a sporadic task system is schedulable by the
  fixed-priority scheduling algorithm if $\sum_{i=1}^{k} U_i \leq 1$
  and 
  \begin{equation}
    \label{eq:R-D-k}
    \frac{C_k}{D_k} \leq 1-\sum_{i=1}^{k-1}  U_i -
    \frac{\sum_{i=1}^{k-1}  C_i}{D_k} + \frac{\sum_{i=1}^{k-1} U_i (\sum_{\ell=i}^{k-1}  C_{\ell})}{D_k},
  \end{equation}
  where the $k-1$ higher-priority tasks are ordered
  in a non-increasing order of their periods.
\end{corollary}

\subsection{Analytical Comparison of \frameworkkq{} and \frameworkku{}}

The utilization-based worst-case response-time analysis in  
Theorem~\ref{theorem:response-time-sporadic} is analytically tighter  
than the best known result, $R_{k} \leq \frac{C_k+ \sum_{i=1}^{k-1}
  C_i - \sum_{i=1}^{k-1} U_i C_i}{1-\sum_{i=1}^{k-1} U_i}$, by Bini et  
al. \cite{bini2009response}. 
 Lehoczky \cite{DBLP:conf/rtss/Lehoczky90} also provides the total
utilization bound of RM scheduling for arbitrary-deadline systems. The
analysis in \cite{DBLP:conf/rtss/Lehoczky90} is based on the Liu and
Layland analysis \cite{liu1973scheduling}. The resulting utilization
bound is a function of $\Delta=\max_{\tau_i}\{\frac{D_i}{T_i}\}$. When
$\Delta$ is $1$, it is an implicit-deadline system. The utilization
bound in \cite{DBLP:conf/rtss/Lehoczky90} has a closed-form when
$\Delta$ is an integer. However, calculating the utilization bound for
non-integer $\Delta$ is done asymptotically for $k=\infty$ with
complicated analysis.  Bini \cite{DBLP:journals/tc/Bini15} provides a total utilization bound
of RM scheduling, based on the quadratic response time analysis in
\cite{bini2009response}, that works for any arbitrary ratio of
$\max_{\tau_i}\{\frac{D_i}{T_i}\}$. 

For constrained-deadline sporadic task sets, since the same test in
Eq.~\eqref{eq:exact-test-constrained-deadline-2} is used for
constructing Definition~\ref{def:kpoints-k2u} and
Definition~\ref{def:kpoints}, the result (with respect to the
conditions in Theorem~\ref{theorem:sporadic-general-k2u},
Corollary~\ref{corollary-rm-k2u},
Theorem~\ref{theorem:response-time-sporadic}, and
Corollary~\ref{corollary-rm}) by using \frameworkku{} is superior to
that by using \frameworkkq{}. The speedup factor of the test in
Eq.~\eqref{eq:schedulability-sporadic-k2u-any-a} in
Theorem~\ref{theorem:sporadic-arbitrary-k2u} has been proved to be
$1.76322$, which is also better than that in
Eq.~\eqref{eq:schedulability-sporadic-any-a} in
Theorem~\ref{theorem:schedulability-sporadic-arbitrary}.\footnote{The
 speedup factor for the schedulability test by using Eq.~\eqref{eq:schedulability-sporadic-any-a} is $2$. This
  is obtained by ignoring the last term in the right-hand-side of
  Eq.~\eqref{eq:schedulability-sporadic-any-a}. Since this is not
  analytically superior, the analysis was not shown in \cite{DBLP:journals/corr/abs-kRTA}.} 
However, the quadratic bound in
Eq.~\eqref{eq:schedulability-sporadic-any-a} can be better than the
hyperbolic bound in Eq.~\eqref{eq:schedulability-sporadic-k2u-any-a}, as demonstrated
in the evaluations.

For arbitrary-deadline sporadic task sets, two different
tests are applied: one comes from
Eq.~\eqref{eq:sufficient-test-arbitrary-deadline} for constructing
Theorem~\ref{theorem:sporadic-arbitrary-k2u} and
Theorem~\ref{theorem:schedulability-sporadic-arbitrary} and another
comes from Eqs.~\eqref{eq:sporadic-arbitrary-objective-k} and
\eqref{eq:sporadic-arbitrary-objective-k2} for construction
Theorem~\ref{theorem:response-time-sporadic} and
Corollary~\ref{corollary:arbitrary-response-schedulability}. It should
be clear that the test from
Eqs.~\eqref{eq:sporadic-arbitrary-objective-k} and
\eqref{eq:sporadic-arbitrary-objective-k2} is tighter than that from
Eq.~\eqref{eq:sufficient-test-arbitrary-deadline}. Therefore, these
results are not analytically comparable.

Note that we can also use Lemma~\ref{lemma:framework-general-k2u} by
defining the values of $\alpha_i$ and $\beta_i$ for each task $\tau_i$
in $hp_1(\tau_k)$ precisely to make the hyperbolic bound in
Eq.~\eqref{eq:schedulability-sporadic-k2u-any-a} and
Eq.~\eqref{eq:schedulability-sporadic-arbitrary-k2u-any-a} more
precisely. Their performance will be provided in the evaluation results.

\subsection{Simulation Environment}
\label{subsec:simulation-env}
The rest of this section presents our evaluation results for the above tests. 
We generated a set of sporadic tasks. The cardinality of the task set was $10$.
The UUniFast-Discard method~\cite{davis2011improved} was adopted to generate a set of utilization values with the given goal.
We  used the approach suggested by Davis et
al.~\cite{davis2008efficient}
to generate the task periods according
to a uniform distribution in the range of the logarithm of the task periods (i.e.,
log-uniform distribution).  
The order of magnitude $p$ to control the period values between the largest and smallest periods is parameterized in evaluations, (e.g., $1-10ms$ for $p=1$, $1-100ms$ for $p=2$, etc.). 
We evaluate these tests in uniprocessor systems with $p \in [1,2,3]$.
The priority ordering of the tasks is assigned according to deadline-monotonic (DM) scheduling. The execution time was set accordingly, i.e., $C_{i}=T_iU_i$.

\begin{figure*}[t]
   \centering
  \includegraphics[width=\textwidth]{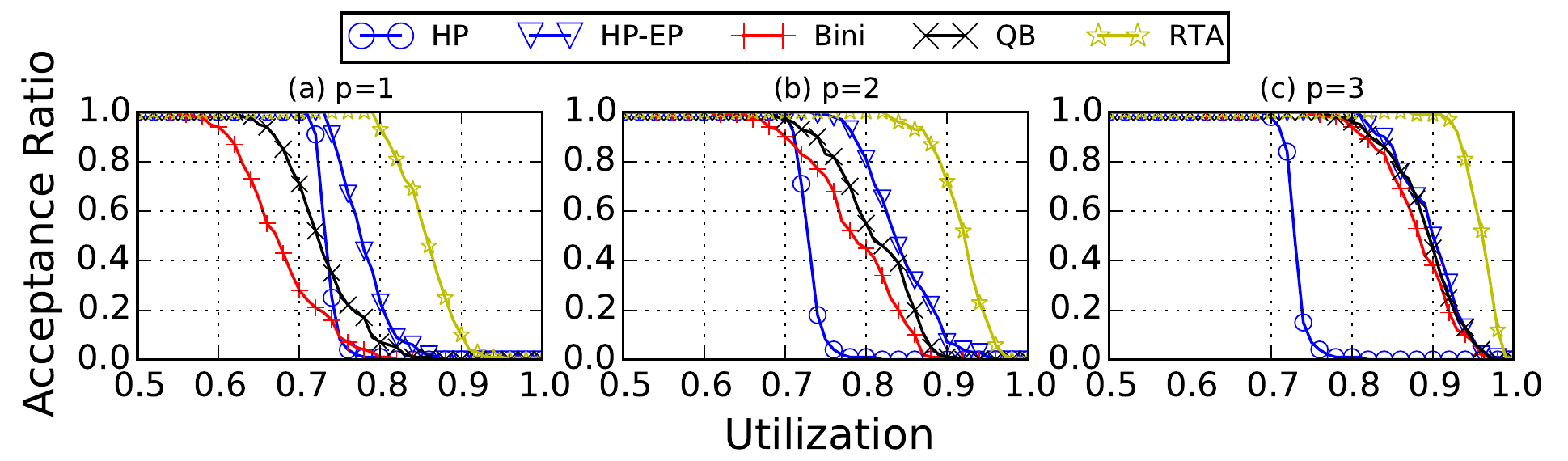}
  \caption{Performance evaluation on uniprocessor systems in terms of acceptance ratio for constrained-deadline uniprocessor systems where $\frac{D_i}{T_i}\in[0.8,1]$.}
    \label{fig:uni-1}
\end{figure*}

The metric to compare results is to measure the \emph{acceptance ratio} of the above tests with respect to a given task set utilization. 
We generate 100 task sets for each utilization level.
The acceptance ratio of a level is said to be the number of task sets that are schedulable under the schedulability test divided by the number of task sets for this level, i.e., 100.

\subsection{Evaluation for Constrained Deadline Systems} 
Task relative deadlines were uniformly drawn from the interval $[0.8T_i,T_i]$.  
The evaluated  tests are as follows:
\begin{itemize}
\item \emph{RTA}: the exact response time test by Lehoczky et al. \cite{DBLP:conf/rtss/LehoczkySD89}.
\item \emph{Bini}: the linear-time response time bound by Bini et al.~\cite{bini2009response}.
\item \emph{HP} (from \frameworkku{}): Eq.~\eqref{eq:schedulability-sporadic-k2u-any-a} in Theorem~\ref{theorem:sporadic-general-k2u} in this report.
\item \emph{HP-EP} (from \frameworkku{}): using Lemma~\ref{lemma:framework-general-k2u}
  (with a more precise extreme point) by defining the values of
  $\alpha_i$ and $\beta_i$ for each task $\tau_i$ in $hp_1(\tau_k)$
  precisely in this report. This improves HP.
\item \emph{QB} (from \frameworkkq{}): Eq.~\eqref{eq:schedulability-sporadic-any-a} in Theorem~\ref{theorem:sporadic-general} in this report.
\end{itemize}

\vspace{0.1in}
{\noindent \bf Results.}
Figure~\ref{fig:uni-1} shows that the performance of the above tests in terms of acceptance ratios, for three different settings of $p$.
The tests by HP-EP, Bini, QB, and RTA are sensitive to $p$: the larger the value of $p$ is, the more the test sets they admit. In the case of $p=1$, the test by Bini (the QB test, respectively) can admit all task sets with their total utilizations of up to $55\%$ ($60\%$, respectively), and its performance starts to decline at utilization $55\%$ ($60\%$, respectively). On the other hand, the tests by HP and HP-EP can fully accept a task set with around $15\%$ more utilizations, but acceptance ratio of HP drops sharply and becomes completely ineffective at utilization $76\%$. 

In the case of $p=1$, we can also see that test HP derived from
\frameworkku{} and test QB derived from \frameworkkq{} are
incomparable. HP itself becomes pessimistic since we do not take the
different values of $\alpha_i$ to have more precise tests, whereas
HP-EP is more precise. In general, for
uniprocessor constrained-deadline task systems, we can observe that
HP-EP outperforms the other polynomial-time tests.  Due to the
analytical dominance, we also see that the QB test dominates the test
by Bini.

\subsection{Evaluation for Arbitrary-Deadline Systems}

Task relative deadlines were uniformly drawn from the interval $[T_i,2T_i]$.  
The tests evaluated are shown as follows:
\begin{itemize}
\item \emph{RTA}: the exact response time test by Lehoczky\cite{DBLP:conf/rtss/Lehoczky90}.
\item \emph{Bini}: the linear-time response time bound by Bini et al.~\cite{bini2009response}.
\item \emph{HP-Busy} (from \frameworkku{}): Eq.~\eqref{eq:schedulability-sporadic-arbitrary-k2u-any-a} from Theorem~\ref{theorem:sporadic-arbitrary-k2u}  in this report
\item \emph{HP-EP} (from \frameworkku{}): using Lemma~\ref{lemma:framework-general-k2u}
  (with a more precise extreme point) by defining the values of
  $\alpha_i$ and $\beta_i$ for each task $\tau_i$ in $hp_1(\tau_k)$
  precisely in this report. This improves HP-Busy. 
\item \emph{QB-Busy} (from \frameworkkq{}): Eq.~\eqref{eq:schedulability-sporadic-any-arbitrary-a} in Theorem~\ref{theorem:schedulability-sporadic-arbitrary} in this report.
\item \emph{QB-Response} (from \frameworkkq{}): Eq.~\eqref{eq:R-D-k} in Corollary~\ref{corollary:arbitrary-response-schedulability} in this report.
\end{itemize}

\begin{figure*}[t]
  \centering
   \includegraphics[width=\textwidth]{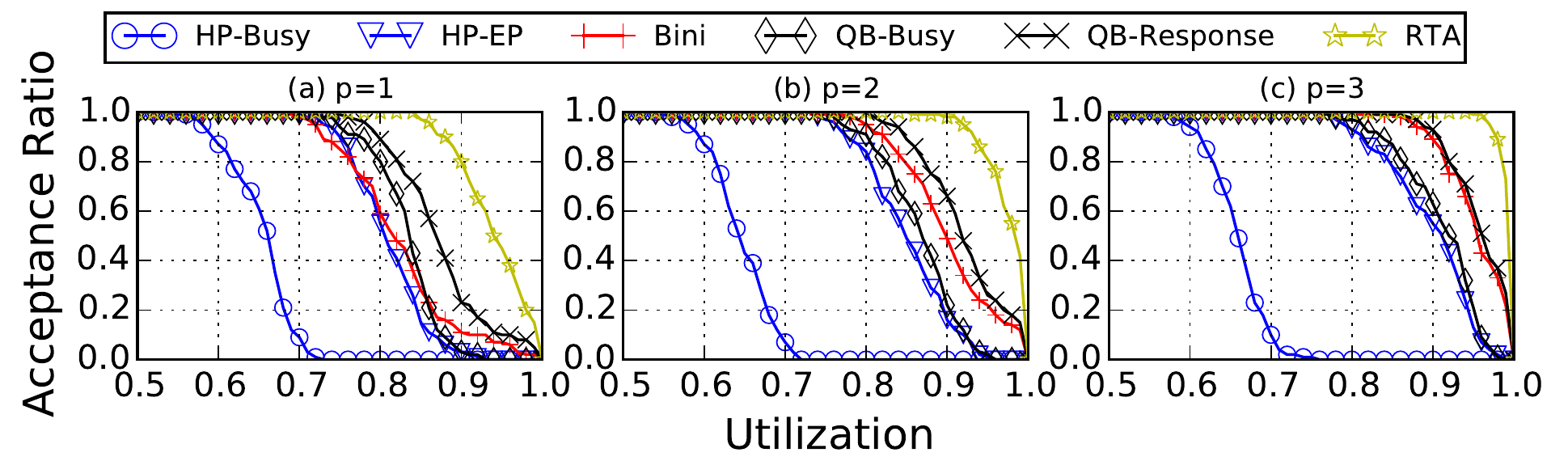}
  \caption{Performance evaluation on uniprocessor systems in terms of acceptance ratio for arbitrary-deadline uniprocessor systems where $\frac{D_i}{T_i}\in[1,2]$}
    \label{fig:uni-2}
\end{figure*}

{\noindent \bf Results.}  Figure~\ref{fig:uni-2} compares the
performance on arbitrary-deadline uniprocessor system where
$\frac{D_i}{T_i}\in [1,2]$.  
Analytically, we know that test by QB is superior to that by Bini,
which is the best-known test for arbitrary-deadline uniprocessor
systems.  The results shown in Figure~~\ref{fig:uni-2} also support
such dominance.

In the case of $p=1$, the acceptance ratio by Bini decreases steadily from utilization $68\%$ to $95\%$.
On the other hand, the number of task sets accepted by QB-Response starts to decrease at utilization $75\%$. Test QB-Response is able to admit more task tests from utilization $68\%$ to $85\%$, compared to test Bini. With utilization more $85\%$, Bini performs better than QB-Busy. In the other cases, Bini outperforms QB-Busy.

For arbitrary-deadline systems, since the test in
Eq.~\eqref{eq:sufficient-test-arbitrary-deadline} is too pessimistic
by checking whether the busy-window length is no more than $D_k$,
HP-Busy, HP-EP, and QB-Busy do not perform very well. In the above
experimental results, the quadratic forms by using \frameworkkq{} are
better than the hyperbolic forms by using \frameworkku{} in such
cases. This is due to the fact that these two tests start from
different pseudo-polynomial time tests.

%
%

\section{Global RM Scheduling}
\label{sec:global-sporadic}

This section demonstrates the two frameworks 
for multiprocessor global fixed-priority scheduling. We consider that
the system has $M$ identical processors. For global fixed-priority
scheduling, there is a global queue and a global scheduler to dispatch
the jobs. We demonstrate the applicability for implicit-deadline
sporadic systems under global RM.

Unfortunately, unlike uniprocessor systems, up to now, there is no
exact schedulability test to verify whether task $\tau_k$ is schedulable
by global RM.  Therefore, existing schedulability
tests (in pseudo-polynomial time or exponential time) are only
sufficient tests. We will use three different tests for demonstrating
the use of the \frameworkku{} and \frameworkkq{} frameworks and compare their results.

One way to quantify the quality of the resulting schedulability test
is to use the \emph{capacity augmentation factor} \cite{Li:ECRTS14}.
Suppose that the test is to verify whether the total utilization
$\sum_{\tau_i} \frac{U_i}{M} \leq \frac{1}{b}$
 and the maximum utilization $\max_{\tau_i}
U_i  \leq \frac{1}{b}$. Such a factor $b$ has
been recently named as a \emph{capacity augmentation factor}
\cite{Li:ECRTS14}. 

We only consider testing the schedulability of task $\tau_k$ under
global RM, where $k > M$. For $k \leq M$, the global RM scheduling
guarantees the schedulability of task $\tau_k$ if $U_k \leq
1$. Without loss of generality, we limit our presentation to the
case that $T_i < T_k$ for
$i=1,2,\ldots,k-1$, for the simplicity of presentation.

\subsection{Adopted Pseudo-Polynomial-Time and Exponential-Time Tests}

We now present three different tests that require
pseudo-polynomial-time or exponential-time complexity.

\noindent{\bf Greedy-Carry-In}:  The first one is based on a simple observation to carry-in a job for
each of the higher-priority tasks in the window of interest \cite{DBLP:conf/rtss/GuanSYY09}. The
following time-demand function $W_i(t)$ can be used for a simple
sufficient schedulability test:
\begin{equation}
  \label{eq:W_i-multiprocessor}
W_i(t) = \left(\ceiling{\frac{t}{T_i}}-1\right)C_i + 2C_i.  
\end{equation}
That is, we allow the first release of task $\tau_i$ to be inflated by
a factor $2$, whereas the other jobs of task $\tau_i$ have the same
execution time $C_i$.  
Therefore, task $\tau_k$ is schedulable under global RM on $M$
identical processors, if
{\small  \begin{equation}
    \label{eq:gDM-sufficient}
\exists t \mbox{ with } 0 < t \leq T_k {\;\; and \;\;} C_k
+\sum_{\tau_i \in hp(\tau_k)}
\frac{W_i(t)}{M} \leq t, 
  \end{equation}
}
 
\noindent{\bf Bounded-Carry-In}:  
The second test is based on the observation by Guan et
al. \cite{DBLP:conf/rtss/GuanSYY09} that we only have to consider
$M-1$ tasks with carry-in jobs, for constrained-deadline task sets.
 For implicit-deadline
task sets, this means that we only need to set $\alpha_i$ of some
tasks to $\frac{2}{M}$, rather than all the $k-1$ tasks in
Eq.~\eqref{eq:W_i-multiprocessor}. More precisely, we can define two
different time-demand functions, depending on whether task $\tau_i$ is
with a carry-in job or not:\footnote{This is an over-approximation of the linear function used by Guan et al. \cite{DBLP:conf/rtss/GuanSYY09}.}
\begin{equation}
  \label{eq:W_i-carryin}
W_i^{carry}(t) =
\begin{cases}
  C_i & 0 < t < C_i\\
  C_i + \ceiling{\frac{t-C_i}{T_i}}C_i & otherwise,
\end{cases}
\end{equation}
and
\begin{equation}
  \label{eq:W_i-normal}
W_i^{normal}(t) = \ceiling{\frac{t}{T_i}}C_i.
\end{equation}
Moreover, we can further over-approximate $W_i^{carry}(t)$, since  $W_i^{carry}(t) \leq W_i^{normal}(t)+C_i$. Therefore, a sufficient schedulability test for testing task $\tau_k$ with $k > M$ for global RM is to verify whether 
\begin{equation}
  \label{eq:grm-multiprocessor-M-1-carryin}
\exists 0 < t \leq T_k, C_k + \frac{(\sum_{\tau_i \in {\bf T}'} C_i) +  (\sum_{i=1}^{k-1}W_i^{normal}(t)) }{M} \leq t.  
\end{equation}
for all ${\bf T}' \subseteq hp(\tau_k)$ with $|{\bf T}'| = M-1$. It is
not necessary to enumerate all ${\bf T}'$ with $|{\bf T}'| = M-1$ if
 we can construct the task set ${\bf T}' \subseteq hp(\tau_k)$
with the maximum $\sum_{\tau_i \in {\bf T}'} C_i$.

\noindent{\bf Forced-Forward}: The third one is based on a
reformulation of the forced-forward approach by Baruah et
al. \cite{DBLP:journals/rts/BaruahBMS10}. This is the reformulation in
\cite{DBLP:journals/corr/abs-kRTA} based on a simple revision of the
forced-forward algorithm in \cite{DBLP:journals/rts/BaruahBMS10}.  Let $U_k^{\max}$ be
$\max_{j=1}^{k} \{U_j\}$.  As shown and proved in
\cite{DBLP:journals/corr/abs-kRTA}, task $\tau_k$ in a
sporadic task system with implicit deadlines is schedulable by a
global RM on
$M$ processors if
\begin{align}
  \forall y \geq 0, \forall 0 \leq \omega_i \leq T_i, \forall \tau_i \in hp(\tau_k), \exists t \mbox{ with } 0 < t \leq T_k+y  {\;\; \;\;} \nonumber\\
  U_k^{\max} \cdot(T_k+y) +
 \frac{ \sum_{i=1}^{k-1} \omega_i\cdot U_i+ \ceiling{\frac{t-\omega_i }{T_i}} C_i
}{M}\leq t. \label{eq:test-forced-forward}
\end{align}

The schedulability condition in
Eq.~\eqref{eq:test-forced-forward} requires to test all possible $y
\geq 0$ and all possible settings of $0 \leq \omega_i \leq T_i$ for
the higher priority tasks $\tau_i$ with $i=1,2,\ldots,k-1$.
Therefore, it needs exponential time (for all the possible
combinations of $\omega_i$).

\subsection{Polynomial-Time Tests by \frameworkku{}}

We now demonstrate how the \frameworkku{} framework can be adopted.

\noindent\underline{\bf Based on Greedy-Carry-In}: 
Such a case is pretty clear by setting $0 < \alpha_i \leq \frac{2}{M}$
and $0 < \beta_i \leq \frac{1}{M}$ in Definition~\ref{def:kpoints-k2u}
for task $\tau_i \in hp(\tau_k)$. Therefore, by using
Lemma~\ref{lemma:framework-constrained-k2u} and
Lemma~\ref{lemma:framework-totalU-exclusive-k2u}, we have the
following theorem.

\begin{theorem}
\label{thm:multiprocessor-GRM}
Task $\tau_k$ in a sporadic implicit-deadline task system is
schedulable by global RM on $M$ processors if
\begin{equation}
\label{eq:schedulability-GRM}
 (\frac{C_k}{T_k}+2)\prod_{\tau_i \in hp(\tau_k)} (\frac{U_i}{M} + 1)\leq 3,
\end{equation}
or
\begin{equation}
\label{eq:schedulability-augmentation-GRM}
\sum_{\tau_i \in hp(\tau_k) } \frac{U_i}{M}\leq \ln{\frac{3}{\frac{C_k}{T_k}+2}}.
\end{equation}
\end{theorem}

\noindent\underline{\bf Based on Bounded-Carry-In}: There are two ways
to use \frameworkku{}. In the first case, we consider that $C_i$ for
task $\tau_i$ in $hp(\tau_k)$ is known. For such a case, we simply
have to put the $M-1$ higher-priority tasks with the largest execution
times into ${\bf T}'$. This can be imagined as if we increase the
execution time of task $\tau_k$ from $C_k$ to $C_k' = C_k +
\frac{\sum_{\tau_i \in {\bf T}'} C_i}{M}$. Therefore, we still have $0
< \alpha_i \leq \frac{1}{M}$ and $0 < \beta_i \leq \frac{1}{M}$ for
$\tau_i \in hp(\tau_k)$. Therefore, by using
Lemma~\ref{lemma:framework-constrained-k2u} and
Lemma~\ref{lemma:framework-totalU-exclusive-k2u}, we have the
following theorem:

\begin{theorem}
\label{thm:multiprocessor-GRM-M-1-V0}
Task $\tau_k$ in a sporadic implicit-deadline task system is
schedulable by global RM on $M$ processors if
\begin{equation}
\label{eq:schedulability-GRM-M-1-V0}
 (\frac{C_k'}{T_k}+1)\prod_{\tau_i \in hp(\tau_k)} (\frac{U_i}{M} + 1)\leq 2,
\end{equation}
or
\begin{equation}
\label{eq:schedulability-augmentation-GRM-M-1-V0}
\sum_{\tau_i \in hp(\tau_k) } \frac{U_i}{M}\leq \ln{\frac{2}{\frac{C_k'}{T_k}+1}},
\end{equation}
where $C_k' = C_k + \frac{\sum_{\tau_i \in {\bf T}'} C_i}{M}$.
\end{theorem}

In the second case, if only the task utilizations are given, we are not sure which tasks
should be put into the carry-in task set ${\bf T}'$. That is, if we
are testing the worst-case period assignments of the higher-priority
tasks in $hp(\tau_k)$, we need to enumerate ${\bf T}'$.  Nevertheless,
if ${\bf T}'$ with $|{\bf T}'|=M-1$ is specified, the translation to
the \frameworkku{} framework is as follows: (1) the parameters are $0
< \alpha_i \leq \frac{2}{M}$ and $0 < \beta_i \leq \frac{1}{M}$ by
using Eq.~\eqref{eq:grm-multiprocessor-M-1-carryin} if $\tau_i$ is in
${\bf T}'$, and (2) the parameters are $\alpha_i = \frac{1}{M}$ and $0
< \beta_i \leq \frac{1}{M}$ by using
Eq.~\eqref{eq:grm-multiprocessor-M-1-carryin} if $\tau_i$ is not in
${\bf T}'$. It may seem at first glance that we have to check all
possible permutations of ${\bf T}'$. Fortunately, with the analysis in
\cite{DBLP:journals/corr/abs-1501.07084,DBLP:conf/rtss/ChenHL15}, the worst permutation of
${\bf T}'$ is to the $M-1$ higher-priority tasks with the largest
utilization into ${\bf T}'$. This leads to the following theorem by
extending Lemma~\ref{lemma:framework-general-k2u}.

\begin{theorem}
\label{thm:multiprocessor-GRM-M-1-carry}
Task $\tau_k$ in a sporadic implicit-deadline task system is
schedulable by global RM on $M$ processors if
\begin{equation}
\label{eq:schedulability-GRM-M-1-carry}
0 < U_k\leq 1 -  \sum_{i=1}^{k-1}  \frac{U_i(\alpha_i
  +\frac{1}{M})}{\prod_{j=i}^{k-1} (\frac{1}{M} U_j + 1)},
\end{equation}
by indexing the $k-1$ higher-priority tasks in a non-decreasing order of $U_i$ and assigning $\alpha_1, \alpha_2, \ldots, \alpha_{k-M}$ to $\frac{1}{M}$ and $\alpha_{k-M+1}, \alpha_{k-M+2}, \ldots, \alpha_{k-1}$ to $\frac{2}{M}$.
\end{theorem}

\noindent\underline{\bf Based on Forced-Forward}:
Formulating the test in Eq.~\eqref{eq:test-forced-forward} into the
\frameworkku{} framework is problematic. Suppose that $T_k$ is
$1$. Assume that $y$ is set to $0$, $\omega_i$ is set to $0.1$,
and $T_i$ is set to $0.92$. Under the above setting, $t_i$ is $0.1$,
and $\alpha_i$ is $1$, $\beta_i$ is $9.2$. In fact, we even cannot
safely set $\beta_i$ to any possible value except $\infty$ if $t_i$ is
small enough. Therefore, constructing parameters based on Definition
\ref{def:kpoints-k2u} is not possible (or non-trivial).

\subsection{Polynomial-Time Tests by \frameworkkq{}}

\noindent\underline{\bf Based on Greedy-Carry-In}: This is
possible by setting $0 < \alpha_i \leq \frac{2}{M}$ and $0 < \beta_i
\leq \frac{1}{M}$
and applying
Lemma~\ref{lemma:framework-general-schedulability}. However, since the
results are not superior to the one with bounded-carry-in, we omit it.

\noindent\underline{\bf Based on Bounded-Carry-In}:
To use \frameworkkq{}, we are certain about which tasks should be put
into the carry-in task set ${\bf T}'$ by assuming that $C_i$ and $T_i$
are both given. That is, we simply have to put the $M-1$
higher-priority tasks with the largest execution times into ${\bf
  T}'$. This can be imagined as if we increase the execution time of
task $\tau_k$ from $C_k$ to $C_k' = C_k + \frac{\sum_{\tau_i \in {\bf
      T}'} C_i}{M}$.

This leads to the following theorem by using
Lemma~\ref{lemma:framework-general-schedulability}. 

\begin{theorem}
\label{thm:multiprocessor-GRM-M-1-carry-k2q}
Task $\tau_k$ in a sporadic implicit-deadline task system is
schedulable by global RM on $M$ processors if $\sum_{i=1}^{k-1}
  C_i \leq M T_k$ and
\begin{equation}\small
\label{eq:schedulability-GRM-M-1-carry-k2q}
 U_k \leq 1 -   \frac{\sum_{\tau_i \in {\bf
      T}'} C_i}{M T_k} -  \sum_{i=1}^{k-1} \frac{U_i}{M} -
\frac{\sum_{i=1}^{k-1} C_i}{M T_k}+\frac{\sum_{i=1}^{k-1} ( U_i
  \sum_{\ell=i}^{k-1} C_\ell )}{M^2 T_k}.
\end{equation}
by indexing the $k-1$ higher-priority tasks in a non-decreasing order
of $(\ceiling{\frac{T_k}{T_i}}-1)T_i$ and by putting the $M-1$
higher-priority tasks with the largest execution times into ${\bf
  T}'$.
\end{theorem}

We can of course revise the statement in
Theorem~\ref{thm:multiprocessor-GRM-M-1-carry-k2q} by adopting
Lemma~\ref{lemma:framework-constrained-schedulability} and
Lemma~\ref{lemma:framework-totalU-exclusive} to construct
schedulability tests by using only task utilizations.

\noindent\underline{\bf Based on Forced-Forward}: 
We present the corresponding polynomial-time schedulability tests for
global fixed-priority scheduling. By using the forced-forward test,
we can adopt the \frameworkkq{} framework by setting
$\alpha_i=\frac{1}{M}$ and $\beta_i = \frac{1}{M}$. Due to the fact
that $T_i \leq T_k$ for any task $\tau_i \in hp(\tau_k)$, i.e., $C_i
\leq U_i T_k$, under global RM, we can reach the following theorems and
corollary, where the proofs are in \cite{DBLP:journals/corr/abs-kRTA}.

 \begin{theorem}
  \label{thm:multiprocessor-grm-sporadic-k2q-tight}
Let $U_k^{\max}$ be $\max_{j=1}^{k} U_j$.  
Task $\tau_k$ in a sporadic task system with implicit deadlines  is schedulable by global RM on
$M$ processors if
\begin{equation}
\label{eq:schedulability-GRM-k2q-tight}
U_k^{\max} \leq 1 -  \sum_{i=1}^{k-1} \frac{U_i}{M} -
\frac{\sum_{i=1}^{k-1} C_i}{M T_k} +\frac{ \sum_{i=1}^{k-1} U_i
  (\sum_{\ell=i}^{k-1} C_\ell) )}{M^2 T_k},
\end{equation}
by ordering the $k-1$ higher-priority tasks in a non-increasing order of $T_i$.
\end{theorem}

\begin{theorem}
  \label{thm:multiprocessor-grm-sporadic-tight}
Let $U_k^{\max}$ be $\max_{j=1}^{k} U_j$.  
Task $\tau_k$ in a sporadic task system with implicit deadlines  is schedulable by global RM on
$M$ processors if
\begin{equation}
\label{eq:schedulability-GRM-tight}
U_k^{\max} \leq 1 - \frac{2}{M}\sum_{i=1}^{k-1} U_i + \frac{0.5}{M^2}\left((\sum_{i=1}^{k-1} U_i)^2 +(\sum_{i=1}^{k-1} U_i^2)\right)
\end{equation}
or
\begin{equation}
\label{eq:schedulability-GRM-tight-ubound}
\frac{\sum_{j=1}^{k-1} U_j}{M}\leq \left(\frac{k-1}{k}\right)\left(
  2-\sqrt{2+ 2U_k^{\max}\frac{k}{k-1}}\right).
\end{equation}
\end{theorem}

\begin{corollary}
  \label{col:grm-tight}
  The capacity augmentation factor of global RM for a sporadic system
  with implicit deadlines is $\frac{3+\sqrt{7}}{2}\approx 2.823$.
\end{corollary}

\subsection{Analytical Comparison of \frameworkkq{} and \frameworkku{}}

\begin{figure*}[t]
  \centering
  \includegraphics[width=\textwidth]{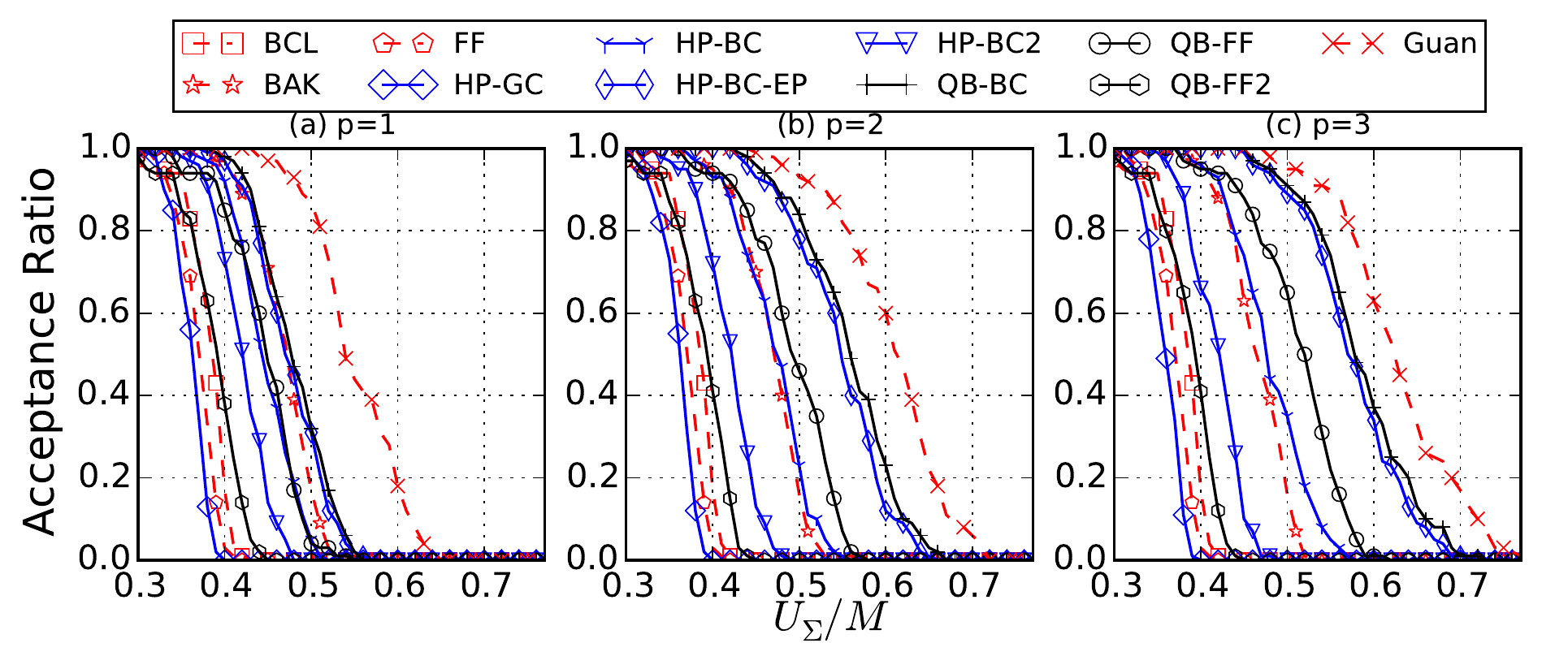}
  \caption{Acceptance ratio comparison on implicit-deadline 8 multiprocessor systems.}
  \label{fig:mul-2}
\end{figure*}

The utilization-based worst-case response-time analysis in
Theorem~\ref{thm:multiprocessor-grm-sporadic-tight} and
Corollary~\ref{col:grm-tight} is analytically tighter than the best
known result by Bertogna et
al. \cite{bertogna2006new} with linear-time tests. Moreover, our
polynomial-time schedulability test extended to handle
deadline-monotonic scheduling for constrained-deadline task sets based
on the forced-forward analysis in
\cite{DBLP:journals/corr/abs-kRTA} has the same speedup factor as the
best known result in pseudo-polynomial time by Baruah et
al. \cite{DBLP:journals/rts/BaruahBMS10}.

With respect to the capacity augmentation factors, the test derived
from \frameworkkq{} by using the forced-forward approach obtains the
best one, whereas the tests from bounded carry-in are worse.\footnote{They can be easily obtained by setting $0 < \alpha_i \leq \frac{2}{M}$ and $0 < \beta_i \leq \frac{1}{M}$.}
As shown in the above examples, different schedulability tests may
lead to different quality of the schedulability tests. Therefore,
these results are not analytically comparable. We will have to compare
these results in the evaluations.

\subsection{Evaluation Results}
In this section, we conduct experiments using synthesized task sets for evaluating the proposed tests on multiprocessor systems.
We first generated a set of sporadic tasks. The cardinality of the task set was $5$ times the number of processors, e.g., 40 tasks on 8 multiprocessor systems.
The task sets were generated in a similar manner in Section~\ref{subsec:simulation-env}. Tasks' relative deadlines were equal to their periods.

The evaluated tests are as follows:
\begin{itemize}
\item \emph{BCL}: the linear-time test in Theorem 4 in~\cite{bertogna2006new}.
\item \emph{FF}: the pseudo-polynomial-time force-forward (FF) analysis in Eq. (5) in \cite{DBLP:journals/rts/BaruahBMS10}.
\item \emph{BAK}: the $O(n^3)$ test in Theorem 11 in~\cite{baker2006analysis}.
\item \emph{Guan}: the pseudo-polynomial-time response time analysis\cite{DBLP:conf/rtss/GuanSYY09}.
\item \emph{HP-GC} (from \frameworkku{}): Eq.~\eqref{eq:schedulability-GRM} from Theorem~\ref{thm:multiprocessor-GRM} based on greedy carry-in (GC) in this report.
\item \emph{HP-BC} (from \frameworkku{}): Eq.~\eqref{eq:schedulability-GRM-M-1-V0} from Theorem~\ref{thm:multiprocessor-GRM-M-1-V0} based on bounded carry-in (BC) in this report. 
\item \emph{HP-BC-EP} (from \frameworkku{}):  using Lemma~\ref{lemma:framework-general-k2u}
  (with a more precise extreme point) by defining the values of
  $\alpha_i$ and $\beta_i$ for each task $\tau_i$ in $hp(\tau_k)$
  precisely in this report. This improves HP-BC from Theorem~\ref{thm:multiprocessor-GRM-M-1-V0} based on bounded carry-in (BC) in this report. 
\item \emph{HP-BC2} (from \frameworkku{}): Eq.~\eqref{eq:schedulability-GRM-M-1-carry} from Theorem~\ref{thm:multiprocessor-GRM-M-1-carry} based on bounded carry-in (BC) in this report.
\item \emph{QB-BC} (from \frameworkkq{}): Eq.~\eqref{eq:schedulability-GRM-M-1-carry-k2q} in Theorem~\ref{thm:multiprocessor-GRM-M-1-carry-k2q} based on bounded carryin (BC) in this report.
\item \emph{QB-FF} (from \frameworkkq{}): Eq.~\eqref{eq:schedulability-GRM-k2q-tight} from Theorem~\ref{thm:multiprocessor-grm-sporadic-k2q-tight} based on force-forward (FF) in this report.
\item \emph{QB-FF2} (from \frameworkkq{}): Eq.~\eqref{eq:schedulability-GRM-tight} from Theorem~\ref{thm:multiprocessor-grm-sporadic-tight} based on force-forward (FF) in this report.
\end{itemize}

Among the above tests,  BCL, HP-GC, HP-BC2, QB-FF\footnote{We assume that the priority ordering is given. We just have to use the reversed order in Theorem~\ref{thm:multiprocessor-grm-sporadic-k2q-tight}.} and QB-FF2 can be implemented in linear time. Our other tests (HB-BC, HP-BC-EP, QB-BC) require to sort the higher-priority tasks to define the proper last release ordering; therefore, their time complexity is $O(n^2 \log n)$ for a task set with $n$ tasks.

{\noindent \bf Results.}
Figure~\ref{fig:mul-2} depicts the result of the performance comparison. In all the cases, we can see that QB-BC and HP-BC-EP are superior to almost all the other polynomial-time tests.  It may seem that QB-FF is superior to QB-BC when we inspect their schedulability tests. However, the way how we formulated the force-forward algorithm in Eq.~\eqref{eq:test-forced-forward} is also pessimistic by introducing $U_k^{\max}$ instead of just $U_k$. Such inflation from $U_k$ to $U_k^{\max}$ makes the analysis for the worst-case capacity-augmentation factor tighter, but also makes QB-FF with less acceptance ratio when testing tasks with utilization larger than the threshold $\frac{1}{2.84306}$. Therefore, if $U_k^{\max} > U_k + \sum_{\tau_i \in {\bf T}'} \frac{C_i}{M T_k}$, then QB-FF is worse than QB-BC.

The greedy carry-in in HP-GC makes it too pessimistic. However, HP-BC is comparable with BAK. Among the linear-time tests, QB-FF outperforms the others in all the cases. Among the above tests, it is difficult to compare QB-BC and HP-BC-EP, since they perform very closely.  Overall, most of the tests derived by using the two frameworks perform very well with low time complexity.

\section{Conclusion}

This report presents the similarly, difference, and the
characteristics of the \frameworkku{} and \frameworkkq{}
frameworks. These two frameworks have great potential to be used for
deriving polynomial-time schedulability tests \emph{almost automatically}, as soon as
the corresponding parameters in Definitions~\ref{def:kpoints-k2u},
\ref{def:kpoints}, and \ref{def:kpoints-response} can be constructed.
In the past, exponential-time schedulability tests were typically not
recommended and most of time ignored, as this requires very high
complexity. However, by adopting these two frameworks, we have
successfully shown that exponential-time schedulability tests may lead
to good polynomial-time tests by using the \frameworkku{} and
\frameworkkq{} frameworks. Both frameworks are needed and have to be
applied for different cases. With these two frameworks, some difficult
schedulability test and response time analysis problems may be solved
by building a good (or exact) exponential-time test and applying these
two frameworks.

These two frameworks are both useful and needed for different cases
and applications. We have demonstrated their differences in details
and present evaluation results for the schedulability tests derived
from these two frameworks. For some cases, \frameworkku{} is better,
and for some cases, \frameworkkq{} is better.

\begin{spacing}{0.9}
   \noindent{\small {\bf Acknowledgement}: This paper has been
     supported by DFG, as part of the Collaborative Research Center
     SFB876 (http://sfb876.tu-dortmund.de/), and the priority program
     "Dependable Embedded Systems" (SPP 1500 -
     http://spp1500.itec.kit.edu).  }
 \end{spacing}

\footnotesize
\vspace{-0.1in}
\def\IEEEbibitemsep{-0.2pt}
\bibliographystyle{abbrv}
\bibliography{ref,real-time}
\normalsize

\end{document}